\documentclass[a4paper, 12pt]{article}
\usepackage{amsmath} \usepackage{amsfonts} \usepackage{amssymb}\usepackage{latexsym}
\usepackage[english]{babel}
\usepackage{amscd}
\usepackage[dvips,final]{graphics}
\usepackage{locengli,math}
\usepackage{graphicx, Addaletree}

\begin{document}
\begin{center}
\textbf{\LARGE{\textsf{ L-algebras, triplicial-algebras,\\[0.3cm] within an equivalence of categories motivated  \\[0.3cm] by graphs
}}}~\footnote{
{ \it{2000 Mathematics Subject Classification: 05E99, 05C20, 05C05, 16W10, 16W30, 17A30, 17A50, 18D50, 60J99.}}\\
{\it{Key words and phrases: Weighted directed graphs, planar rooted symmetric ternary trees, even trees, (Markov) L-coalgebras, L-algebras, $As^c$-L-bialgebras, triplicial-algebras, $As^c-Trip$-bialgebras, Structure theorems (Cartier-Milnor-Moore), Good triples, L-commutative algebras, $NAP$-algebras.}}\\
\textsf{Email}: ph$\_$ler$\_$math@yahoo.com; \textsf{Mail}: 27, Rue Roux Soignat 69003 Lyon, France.}
\vskip2cm
\large{Philippe {\sc Leroux}}
\end{center}
\vskip2cm
\noindent
{\bf Abstract:}
In a previous work, we gave a coalgebraic framework of directed graphs equipped with weights (or probability vectors) in terms of (Markov) L-coalgebras. They are $K$-vector spaces equipped with two co-operations, $\Delta_M$, $\tilde{\Delta}_M$ verifying,
$$(\tilde{\Delta}_M \otimes id)\Delta_M =(id \otimes \Delta_M)\tilde{\Delta}_M.$$
In this paper, we study the category of L-algebras (dual of L-coalgebras), prove that the free L-algebra on one generator is constructed over rooted planar symmetric ternary trees with odd numbers of nodes and the L-operad is Koszul. We then introduce triplicial-algebras: vector spaces equipped with three associative operations verifying three entanglement relations.
The free triplicial-algebra is computed and turns out to be related to even trees. Via a general structure theorem (\`a la Cartier-Milnor-Moore) proved in Section~4, the category of L-algebras turns out to be equivalent to a much more structured category called connected coassociative triplicial-bialgebras (coproduct linked to operations via  infinitesimal relations), that is the triple of operads $(As, Trip, \textrm{L})$ is good.
Bidirected graphs, related to $NAP$-algebras (L-commutative algebras), are briefly evoked and postponed to another paper.

\section{An algebraic setting over weighted directed graphs}
In the sequel, $K$ will be a characteristic zero field and its unit will always be denoted by $1_K$. The symmetric group over $n$ elements is denoted by $S_n$ and if $\mathcal{P}$ denotes a regular operad, then we write $\mathcal{P}(n)=\mathcal{P}_n \otimes KS_n$, the $K$-vector space of the $n$-ary operations of $\mathcal{P}$, see for instance \cite{lodayren} for notation and basic definitions in operad theory.
The symbol $\circ$ stands for the composition of maps and the notation $v_1 \ldots v_n$ for $v_1 \otimes \ldots \otimes v_n$ with $v_i \in V$, where $V$ is a $K$-vector space.

\noindent
In \cite{Coa}, we introduced a coalgebraic framework to code any weighted directed graphs which are row and locally finite. This coding leads to the so-called L-coalgebras setting.
Recall that a \textit{$L$-coalgebra} $(L, \Delta, \tilde{\Delta})$ is a $K$-vector space equipped with a right co-operation $\Delta: L \xrightarrow{} L^{\otimes 2}$ and a left co-operation $\tilde{\Delta}: L \xrightarrow{} L^{\otimes 2}$,
verifying what we call now
the {\sf{entanglement relation}}:
$$ (\tilde{\Delta} \otimes id)\Delta = (id \otimes \Delta)\tilde{\Delta}.$$
A $L$-coalgebra may have two partial counits. The right counit $\epsilon: L \xrightarrow{} K$  verifying
$ (id \otimes \epsilon)\Delta = id$ and the left counit $\tilde{\epsilon}: L \xrightarrow{} K$
verifying,
$ ( \tilde{\epsilon} \otimes id)\tilde{\Delta} = id.$
It has been proved in \cite{Coa} that directed graphs having no source and sink but weighted by probability vectors yield
(Markov) L-coalgebras with same counits. The co-operation  $\Delta_M$ codes the future of a given vertex:
$\Delta_M(\textrm{Present}):= \textrm{Present} \otimes \textrm{Future},$
and the co-operation  $\tilde{\Delta}$ codes its past:
$\tilde{\Delta}_M(\textrm{Present}):= \textrm{Past} \otimes \textrm{Present}.$
The entanglement relation means that Past, Present and Future are related together as expected,
$$ (Past \otimes Present)\otimes Future=
Past \otimes (Present\otimes Future).$$
Let $(A, \cdot)$ be a unital associative algebra. Two convolutions products can be defined over $Hom_K(KG_0, A)$, where $G_0$ is the set of vertices of a given graph:
$ f \prec g := \cdot (f \otimes g) \Delta \ \ \ \textrm{and} \ \ \
f \succ g := \cdot (f \otimes g) \tilde{\Delta}.$
We get for any maps $f,g,h \in Hom_K(KG_0, A)$, what will be also called an entanglement relation:
$$ (f \succ g) \prec h= f \succ (g \prec h).$$
The $K$-vector space $Hom_K(KG_0, A)$ equipped with these two operations turns out to be a so-called L-algebra. Set $\eta: K \hookrightarrow A$, $1_K \mapsto 1_A$. It has a ``unit'' $1 :=\eta \circ \epsilon = \eta \circ \tilde{\epsilon}$, verifying:
$f \prec 1 = f = 1 \succ f,$
if left and right counits are supposed to be equal.
From \cite{Coa}, it has also been proved that bidirected graphs yield to the so-called L-cocommutative coalgebras, that is, $K$-vector spaces equipped with two co-operations verifying the entanglement relations and the following extra condition:
$\Delta= \tau \circ \tilde{\Delta},$
where $\tau$ is the usual flip map. The entanglement relation becomes,
$$ (id \otimes \tau)(\Delta_M \otimes id)\Delta_M=(\Delta_M \otimes id)\Delta_M.$$
Such coalgebras have also been found by M. Livernet \cite{Liv} under the name $NAP$-coalgebras. The case of bidirected graphs is postponned to another paper.

In this paper, we propose a study of the category of L-algebras. In Section 2, we explicit the dual of the L-operad and find the free L-algebra over a given vector space $V$ thanks to rooted planar symmetric ternary trees with odd numbers of nodes coded by words. We prove the existence of an involution on the free L-algebra over $V$ and compute the free L-monoid over a given set. We also propose a new coding for rooted planar binary trees. The L-operad happens to be Koszul, hence generating functions of the L-operad and its dual are inverse one another for the composition of functions. This gives an algebraic interpretation of the sequence $A006013$ from the On Line Encyclopedy of Integer Sequences. In section 3, we discuss bialgebraic versions to extend L-algebras. We first start with
dealing with two unit actions over the free augmented L-algebra
and obtain a coassociative L-morphism. Nevertheless, one of these cases turns out to be too simple.
In Section 4, we show the existence of a general structure theorem, that is a Cartier-Milnor-Moore type theorem or good triples according to J.-L. Loday for the magmatic operad with $n$ operations $[n]-Mag$ equipped with a coassociative coproduct linked to operations via nonunital infinitesimal relations. We show that
entanglement equations yield primitive relations. Dividing out by
such relations yields many good triples, one of them being the second main results of this paper. Indeed, using this theorem, we
prove in Section 5 that the triple of operad $(As, Trip, \textrm{L})$ is good, where the operad $Trip$ is associated with triplicial-algebras, $Trip$-algebras for short. The free $Trip$-algebra over $V$ is computed and turns out to be related to even trees. We then obtain that the category of L-algebras is equivalent to the category of connected coassociative triplicial-bialgebras. In case of walks over a graph, operations of L-algebras are nonassociative and just code the past or the future of the walk. Requiring that Past, Present and Future are ordered leads to a much more structured objects which are the coassociative triplicial-bialgebras! We explicit relations between these two objects via idempotents (Subsection~4.2.) and the universal envelopping functor $U$ (Subsection~5.3.). Reversing time leads to reverse walk over a graph. This is coded through our objects by an involution.
The free commutative $Trip$-algebra over a $K$-vector space is also given. It turns out to be related to commutative algebras and permutative algebras.
Section 6 shows that the L-operad cannot be an anticyclic operad although $[n]-Mag$ is as soon as $n>1$.
We present here a picture of our main results applied in the particular case of weighted directed graphs.
\begin{center}
\includegraphics*[width=16cm]{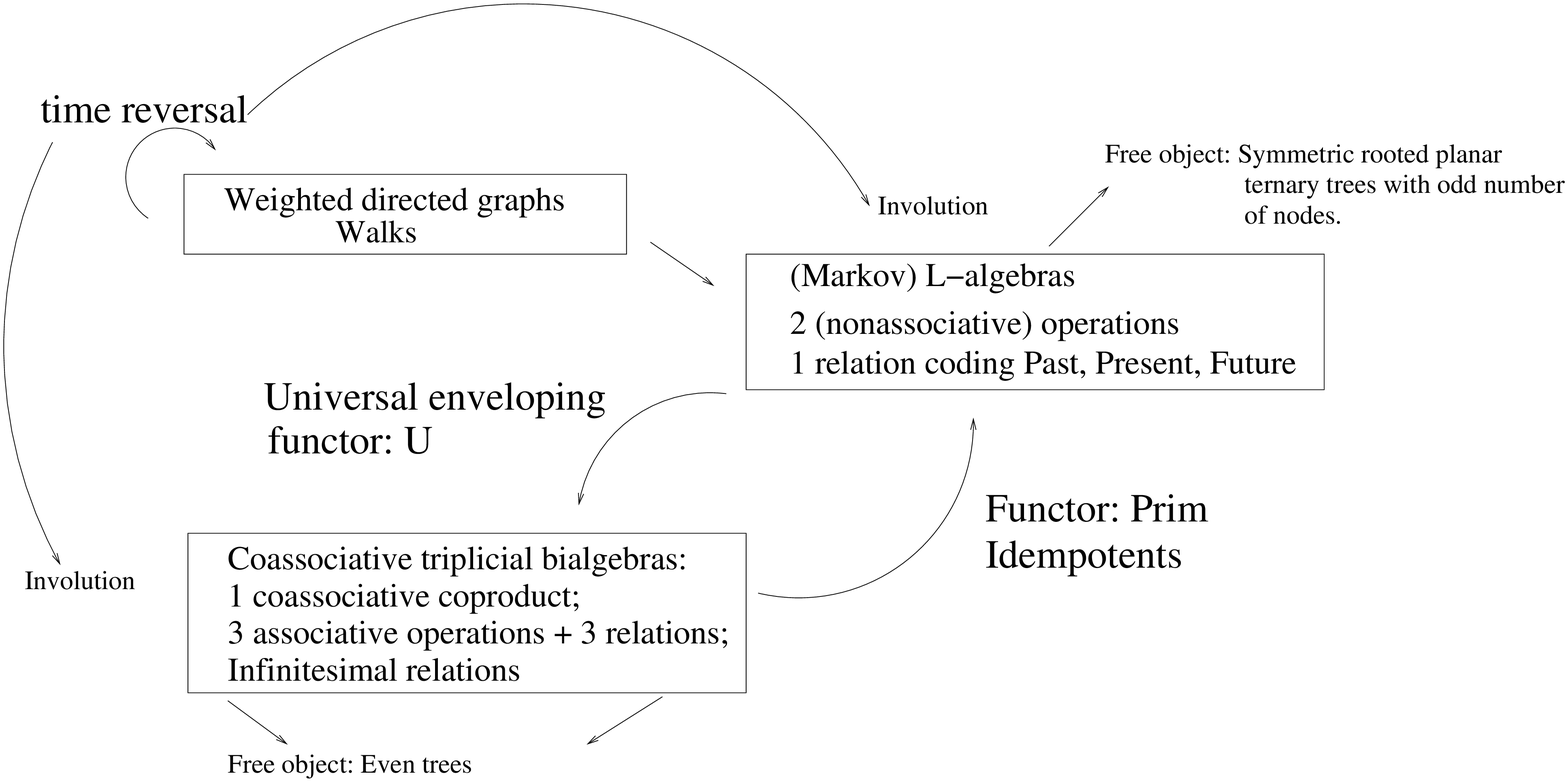}
\end{center}

\section{The free L-algebra and its dual}
\label{freeL}

\begin{defi}{}
A L-algebra is a $K$-vector space $L$ equipped with two binary operations $\prec, \ \succ: L^{\otimes 2} \rightarrow L$ verifying the so-called entanglement relation:
$$ (x \succ y) \prec z = x\succ(y\prec z),$$
for all $x,y,z \in L$.
A L-algebra is said to be involutive if it exists an involution $\dagger: L \rightarrow L$ such that $(x \prec y)^\dagger =  y^\dagger \succ x^\dagger$ and $(x \succ y)^\dagger =  y^\dagger \prec x^\dagger$. The opposite of a L-algebra $L$ is the $K$-vector space $L$ equipped with the operations:
$$ x \prec^{op} y := y \succ x; \ \ \ x \succ^{op}:= y \prec x,$$
for all  $x,y \in L$.
A L-algebra is said to be commutative if it coincides with its opposite. Therefore, a commutative L-algebra is a $K$-vector space equipped with one binary operation $\prec$ verifying:
$$ (x \prec y) \prec z = (x \prec z) \prec y.$$
As mentioned in the introduction, bidirected graphs lead to such structures. They have been also introduced independently by M. Livernet \cite{Liv} under the name $NAP$-algebras.
\end{defi}
\begin{exam}{}
As seen in \cite{Coa}, L-algebras arise from coding weighted directed graphs. But numerous types of algebras are in fact L-algebras. Associative algebras (the two operations coincide with the associative product), magmatic algebras \cite{H} (take the second operation to be zero), dendriform algebras and dialgebras \cite{Loday}, quadri-algebras \cite{AguiarLoday}, ennea-algebras \cite{Lertribax} and all the types of algebras coming from \cite{Ler-polyg}.
In \cite{Coa}, associative L-algebras have been considered. These are L-algebras whose two operations are associative. Such stuctures appear in the previous works of A. Brouder and A. Frabetti \cite{BF} and J.-L. Loday and M. Ronco \cite{LRbruhat}.
They have been renamed in \cite{GB} as duplicial-algebras.
We note also that L-algebras appear in \cite{Chap} without citations to our previous works.
\end{exam}

\noindent
L-algebras give birth to the category \textsf{L-alg} and the so called L-operad which is binary, quadratic and regular.
Consequently, it admits a dual, in the sense of V. Ginzburg and M. Kapranov \cite{GK}, the so called L$^!$-operad. One can check that L$^!$-algebras are defined as follows.

\begin{defi}{}
A L$^!$-algebra is a $K$-vector space $L'$ equipped with two binary operations $\dashv, \ \vdash: L'^{\otimes 2} \rightarrow L'$ such that:
$$(x \vdash y) \dashv z = x \vdash (y \dashv z),$$
\begin{eqnarray*}
(x \vdash y) \vdash z=0, \ &\ & \   x \dashv (y \dashv z) =0,\\
(x \dashv y) \dashv z=0, \ &\ & \   x \vdash (y \vdash z) =0,\\
(x \dashv y) \vdash z=0, \ &\ & \   x \dashv (y \vdash z) =0,
\end{eqnarray*}
hold for all $x,y,z \in L'$.
\end{defi}

\noindent
This category is denoted by \textsf{L$^!$-alg}. Observe that both $\dashv$ and  $\vdash$ are associative and any linear combinations of these two operations as well. If \textsf{As}, \textsf{Dend}, \textsf{Dias},  \textsf{Dup} denote respectively the category of associative algebras, dendriform algebras, dialgebras \cite{Loday}, duplicial-algebras \cite{GB}, then we get the following canonical functors:
$\textsf{L$^!$-alg} \rightarrow \textsf{As}$, $\textsf{L$^!$-alg} \rightarrow \textsf{Dend}$,
$\textsf{L$^!$-alg} \rightarrow \textsf{Dias}$ and
$\textsf{L$^!$-alg} \rightarrow \textsf{Dup}$.

\subsection{The free L$^!$-algebra}
\label{freeL!}
\begin{theo}
Let $V$ be a $K$-vector space. Let $\Psi: (V \oplus K)^{\otimes 3} \rightarrow K$ be the canonical projection.
The free L$^!$-algebra over $V$ is the $K$-vector space,
$$L^!(V):=(V \oplus K)\otimes V \otimes (K \oplus V),$$
equipped with the following operations:
$$ v_1 \otimes v_2 \otimes v_3 \vdash v'_1 \otimes v'_2 \otimes v'_3 :=
\Psi(v_1 \otimes v_3 \otimes v'_1) v_2\otimes v'_2 \otimes v'_3,$$
$$ v_1 \otimes v_2 \otimes v_3 \dashv v'_1 \otimes v'_2 \otimes v'_3 :=
\Psi(v_3 \otimes v'_1 \otimes v'_3) v_1 \otimes v_2 \otimes v'_3.$$
Moreover the generating function of the L$^!$-operad is:
$$ f_{L^!}(x)= x +2x^2 +x^3.$$
\end{theo}
\Proof
Showing that $L^!(V)$ is a L$^!$-algebra is left to the reader. We use the embedding $i: V \rightarrow K \otimes V \otimes K$, $v \mapsto 1 \otimes v \otimes 1$. Let $A$ be a L$^!$-algebra and $f: V \rightarrow A$ be a linear map. We construct its extension $\Phi:L^!(V)\rightarrow A$ as follows:
$$\Phi(v_1 \otimes v_2 \otimes v_3):= f(v_1)\vdash f(v_2) \dashv f(v_3),$$
with the convention that $f(v_1)\vdash $ (resp. $\dashv f(v_3)$) disappears if $v_1=1$ or $v_3=1$. That is $\Phi(v_1 \otimes v_2 \otimes 1):= f(v_1)\vdash f(v_2)$,
$\Phi(1 \otimes v_2 \otimes v_3):= f(v_2) \dashv f(v_3)$ and $\Phi(1 \otimes v_1 \otimes 1):=f(v_1)$.
The map $\Phi$ is a L$^!$-algebra morphism. Indeed, on the one hand,
$\Phi(v_1 \otimes v_2 \otimes v_3 \vdash v'_1 \otimes v'_2 \otimes v'_3)=
\Psi(v_1 \otimes v_3 \otimes v'_1) f(v_2)\otimes f(v'_2) \otimes f(v'_3)$. On the other hand,
$\Phi(v_1 \otimes v_2 \otimes v_3) \vdash \Phi(v'_1 \otimes v'_2 \otimes v'_3)= [f(v_1) \vdash f(v_2) \dashv f(v_3)] \vdash [f(v'_1) \vdash f(v'_2) \dashv f(v'_3)].$
If $v'_1\not=1$, then we get $( \ldots ) \vdash [f(v'_1) \vdash (\ldots )]$ which vanishes. The case $v_1 \not=1$ gives the same result. Suppose now $v_1=v'_1=1$ and $v_3\not=1$, we get $[f(v_2) \dashv f(v_3)]\vdash [f(v'_2) \dashv f(v'_3)]$ which again vanishes. If now
$v_1=v'_1=1=v_3$, then we get
$f(v_2)\vdash (f(v'_2) \dashv f(v'_3))$ showing that $\Phi$ is a morphism for the $\vdash$ operation. The same computation for the other operation shows that $\Phi$ is a L$^!$-algebra morphism which obey $\Phi \circ i=f$. Hence the unicity of $\Phi$ since such a morphism has to coincide on $V$ with $f$.
\eproof
\subsection{Coding rooted planar symmetric ternary trees}
\label{coding}
Before entering the description of the free L-algebra over a $K$-vector space $V$, we need to introduce a combinatorial object, the so-called planar rooted ternary symmetric trees with odd degrees. These rooted trees have internal vertices (or children) with one input, three outputs and
have a reflexive symmetry around the axis passing through the root and its middle child.
Let $W_n$, $n>0$, be the set of the so-called rooted ternary symmetric trees with 2n-1 vertices. Here are $W_1$ and $W_2$:

\begin{center}
\includegraphics*[width=6cm]{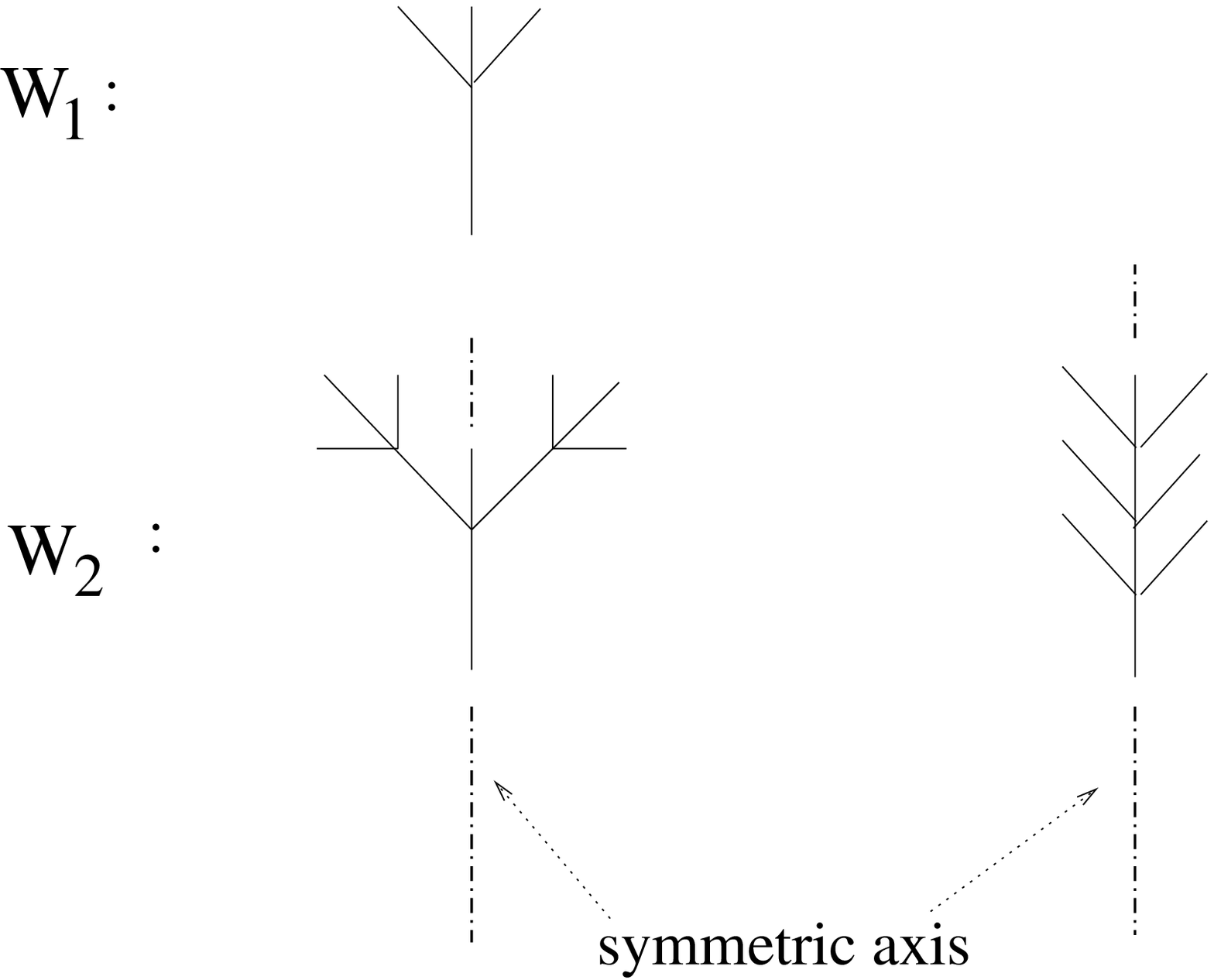}
\end{center}

\noindent
It has been proved by E. Deutsch, S. Feretic and M. Noy \cite{DFN} that the cardinality of $W_n$ is $\frac{1}{n} {3n-2 \choose 2n-1}$ and is registered under the name A006013 in the On-Line Encyclopedia of Integer Sequences. For the purpose of the next subsection, we will use a coding of these trees. In \cite{AMS-D}, E. Deutsch asks for the cardinalities of the following sets:
Let $\mathcal{W}_{2n-1}$, $n>0$, be the set of sequences of integers (called words) of length $2n-1$ such that for each word $\omega:=\omega_1 \omega_2 \ldots \omega_{2n-1} \in \mathcal{W}_{2n-1}$, we have,
\begin{enumerate}
\item{$\omega_1=1;$}
\item{$\omega_i >0$, for all $i \in \{ 1,2,3, \ldots, 2n-1\}$;}
\item{$\omega_i - \omega_{i-1} \in \{1, -1, -3, -5, -7, \ldots \}$, for all $i \in \{2,3,4, \ldots, 2n-1\}$.}
\end{enumerate}
\noindent
For instance,\\
$\mathcal{W}_1 = \{1 \}$, $\mathcal{W}_2 = \{123, 121 \}$ and $\mathcal{W}_3 = \{12345, 12343, 12341, 12323, 12321, 12121, 12123 \}$. It has been proved by D. Callan \cite{AMS-D} that $\mathcal{W}_{2n-1}$ has the same cardinality than $W_n$ (in fact more can be proved if we include planar rooted symmetric ternary trees with even degrees). To enumerate $\mathcal{W}_{2n-1}$, he introduces the following bijection which will be crucial in the next subsection.

\noindent
\textbf{Bijection: from words to lattice paths.}\\
In the sequel, the symbol $\bar{2}$ will stand for the integer $-2$.
We map a word $\omega$ of $\mathcal{W}_{2n-1}$ into a word $l$ of length $3n-2$ with entries $l_i \in \{1, \bar{2} \}$ for all $i$. The word $l$ is built as follows:
\begin{enumerate}
 \item {$l_1=\omega_1$;}
\item{for $2\leq i \leq 2n-1$, $l_i=\omega_i - \omega_{i-1}$;}
\item{if $l_i<0$, then replace it by $(1-l_i)/2$ copies of $\bar{2}$ followed by a 1,}
\item{append enough copies of $\bar{2}$ at the end of the word $l$ such that the sum of the entries of $l$, $\sum_i l_i$, equals 1.}
\end{enumerate}

\noindent
We denote by $\mathcal{L}_n$ the set of words coding trees of $\mathcal{W}_{2n-1}$.
For instance, we get $1 \mapsto 1$, $123 \mapsto 111\bar{2}$ and
$121 \mapsto 11\bar{2}1$ and so on. Consider the lattice $\mathbb{N} \times \mathbb{N}$ and map 1 into the vector $(1;1)$ and $\bar{2}$ into the vector $(2;-2)$. Then, these words describe paths begining at $(0;0)$ and arriving at the point $(4n-3;1)$.
Consequently, we get:
\begin{prop}
\label{carac}
A word $l=l_1l_2 \ldots l_{3n-2} \in \mathcal{L}_n$ if and only if  for all $1\leq k < n$, $\sum_{i=1}^k \ l_i \geq 0$ and $\sum_{i=1}^n \ l_i=1$.
\end{prop}
\noindent
There exists a very simple bijection between words $l$ and the planar symmetric ternary rooted trees with odd degrees found by M. Bousquet and C. Lamathe \cite{BL}.

\subsection{The free L-algebra}
\label{L-alg}

Set $K\mathcal{L}:=\bigoplus_{n>0} K\mathcal{L}_n$.
Let $l \in \mathcal{L}_n$ and $l'\in \mathcal{L}_m$. Define two operations $\succ, \ \prec$ on $K\mathcal{L}$ first by
$$ l \succ l' :=    1 l \bar{2}l',$$
$$ l \prec l' := l1 l' \bar{2},$$
then by bilinearity. Observe that our operations respect the canonical graduation of $K\mathcal{L}$ since,
$$\succ, \ \prec \ : K\mathcal{L}_n \otimes K\mathcal{L}_n \mapsto K\mathcal{L}_{n+m}.$$
\begin{prop}
\label{gene}
The $K$-vector space $K\mathcal{L}$ equipped with the two previous operations is a L-algebra generated by the word 1.
\end{prop}
\Proof
This computation $(l \succ l') \prec l''=    1 l \bar{2} l' 1 l'' \bar{2}=  1 l \bar{2} (l' 1 l'' \bar{2} )       = l \succ (l' \prec l'')$ shows that
$K\mathcal{L}$ is a L-algebra. By hand, one can check that $\mathcal{L}_1, \ \mathcal{L}_2, \ \mathcal{L}_3$ are generated by 1. Fix $n>0$. Suppose this holds up to  $\mathcal{L}_n$. Let $l=1l_1l_2 \ldots l_{3n+1} \in \mathcal{L}_{n+1}$.
First of all, there exists an integer $k_0 \in \{1, \ldots, 3n+1\}$ such that $l_{k_0}=\bar{2}$ and $l':=l_1 \ldots l_{k_0-1}$ is a word. Indeed, if $l_{3n+1}=\bar{2}$, then set $l':=l_1 \ldots l_{3n}$. If $l_{3n+1}=1$, then $l_{3n}=\bar{2}$ because of Proposition~\ref{carac}. Set $l':=l_1 \ldots l_{3n-1}$ to conclude. Let $k_0$ be the smallest integer realizing the previous assertion. Using Proposition~\ref{carac}, if $k_0 < 3n+1$, then there exists a word $l''$ such that
$l= l' \succ l''$, otherwise $l=1 \prec l''$. Therefore, by induction $K\mathcal{L}$ is generated by the word 1 as a L-algebra.
\eproof
\begin{theo}
\label{freeee}
The unique L-algebra map $L(K) \rightarrow (K\mathcal{L}, \prec, \succ)$ sending the generator $x$ of $L(K)$ to the word 1 of $K\mathcal{L}$ is an isomorphism, i.e., $(K\mathcal{L}, \prec, \succ)$ is the free L-algebra on one generator.
\end{theo}
\Proof
Consider the map $i: K \mapsto \textrm{L}(K)$, $1_K \mapsto x$ and the map $f: K \rightarrow K\mathcal{L}$, $1_K \mapsto 1$. As L$(K)$ is the free L-algebra on one generator, there exists a unique L-algebra morphism $\chi: \textrm{L}(K) \rightarrow K\mathcal{L}$ such that $\chi \circ i=f$. For $n=1,2,3$, one can check by hand that the restriction of $\chi$ to $\textrm{L}_n$ into $K\mathcal{L}_n$ is an isomorphism. We suppose this result holds up to an integer $n-1$.
As the L-operad is binary, any monomials of $L_n$ can be written as $X \prec Y$ or $X \succ Y$, with $X \in L_k$ and $Y \in L_{n-k}$ for a $k\in \{1, \ldots, n-1\}$. We get,
$$ \chi(X \prec Y):= \chi(X)1 \chi(Y)\bar{2}, \ \ \chi(X \succ Y):= 1\chi(X) \bar{2} \chi(Y).$$
Let $X' \prec Y'$ and $X \prec Y$ be two monomials of $L_n$, with at least $X\not= X'$ or $Y \not= Y'$. Suppose
$\chi(X' \prec Y')= \chi(X \prec Y)$. Then $\chi(X')1 \chi(Y')\bar{2} =\chi(X)1 \chi(Y)\bar{2}$. If $\chi(X')$ and $\chi(X')$ have the same length then by induction $X=X'$ and thus $Y=Y'$ which is not possible by assumption. Otherwise one of them has a greater length. Suppose  this is $\chi(X')$. Then there exists $u$ made of 1 and $\bar{2}$ such that,
$$\chi(X')= \chi(X)1u,$$
and whose the sum of its entries is -1. Therefore, $\chi(Y)=u \chi(Y')$ which is impossible because of Proposition~\ref{carac}. Hence, $\chi(X' \prec Y')= \chi(X \prec Y)$ does not hold
with our assumption.
Suppose now $\chi(X' \succ Y')= \chi(X \prec Y)$ with at least $X\not= X'$ or $Y \not= Y'$. Then, $1 \chi(X') \bar{2} \chi(Y')= \chi(X) 1 \chi(Y) \bar{2}$. If
$\chi(X')$ and $\chi(X')$ have same length then the word $\chi(Y)$ starts with a $\bar{2}$, which is not possible.
Suppose the length of $\chi(X')$
is greater than $\chi(X)$. Then, set
$$ 1 \chi(X')= \chi(X)1 u,$$
with the sum of the entries equals to 0. Hence, $ \chi(Y)\bar{2}=u \bar{2} \chi(Y')$ which is impossible since the sum of the entries of $u$ and $\bar{2}$ equals -2 and $\chi(Y)$ is a word. Suppose now that the length of $\chi(X)$
is greater than $\chi(X')$. Set,
$$ \chi(X)=1\chi(X')u,$$
with the sum of the entries of $u$ equals to -1. Therefore, $u 1\chi(Y) \bar{2}=\bar{2}\chi(Y')$. Hence, $u$ starts with a $\bar{2}$. Hence, there exists an integer $k$, such that $\chi(Y'):= u_2u_3 \ldots u_k1 \chi(Y)\bar{2}$. But $\sum_{i=2}^k u_i=+1$. Since $\chi(Y')$ is a word,  Proposition~\ref{carac} claims that $u_2u_3 \ldots u_k$ is a word too. By induction, there exists a unique monomial $Z$ of smaller degree such that $\chi(Z):=u_2u_3 \ldots u_k$. But $\chi(X \prec Y)=\chi(X' \succ Y')= \chi(X'\succ Z \prec Y)$. Therefore, $X \prec Y=X' \succ Y'= X'\succ Z \prec Y$.
Hence, for each $n>0$, the restriction of $\chi$ to $L_n$ into $K\mathcal{L}_n$ maps differents monomials into different
words, so
is injective. However Proposition~\ref{gene} show that $\chi$ is surjective, so $\chi$ is an isomorphism.
\eproof

\NB
In the sequel, we denote by $\varpi$ the inverse of $\chi$.
\NB \textbf{[Involution and time reversal]}
There exists a natural involution $\dagger$ over $K\mathcal{L}$ built by induction. As any word of $K\mathcal{L}_n$ is generated uniquely by 1 one defines  $\dagger$ as follows. First of all $1^\dagger=1$, $(l \prec l')^\dagger= l'^\dagger \succ l^\dagger$ and $(l \succ l')^\dagger= l'^\dagger \prec l^\dagger$, by linearity then. This involution is important since reversing time when dealing with walks over a weighted directed graph, viewed as a (Markov) L-algebra, can be coded through this involution.

\noindent
Because the operad $L$ is regular, we get the following result.
\begin{theo}
Let $V$ be a $K$-vector space. Then, the $K$-vector space,
$$\bigoplus_{n>0} K\mathcal{L}_n \otimes V^{\otimes n},$$
equipped with the following binary operations:
$$ (l \otimes v_1\ldots v_n) \prec (l' \otimes v_1\ldots v_{n'})=(l \prec l') \otimes v_1\ldots v_n v_1\ldots v_{n'}, $$
$$ (l \otimes v_1\ldots v_n) \succ (l' \otimes v_1\ldots v_{n'})=(l \succ l') \otimes v_1\ldots v_n  v_1\ldots v_{n'}, $$
is the free L-algebra over $V$. Otherwise stated, the unique L-algebra map $\textrm{L}(V) \rightarrow \bigoplus_{n>0} K\mathcal{L}_n \otimes V^{\otimes n}$
sending $v \in V$ to $1 \otimes v$ is an isomorphism.
Moreover, the generating function of the L-operad is,
\begin{eqnarray*}
f_L(x) &=& \frac{4}{3} \ sin^2(\frac{1}{3} \ asin(\sqrt{\frac{27x}{4}})),\\
&=& \sum_{n>0} \frac{1}{n} {3n-2 \choose 2n-1} x^n =x +2x^2 +7x^3 +30x^4 +\ldots.
\end{eqnarray*}
\end{theo}
\noindent
The next result is a consequence of the theory developed by V. Ginzburg and M. Kapranov \cite{GK}.
\begin{prop}
\label{asso-GK}
Let $(L,\prec, \succ)$ be a L-algebra and $(M, \dashv,\vdash)$ be a L$^!$-algebra.
Then the binary operation $*$ on $M \otimes L$ defined by
$$ (x \otimes a) *(y \otimes b):=(x \dashv y) \otimes (a \prec b) + (x \vdash y)\otimes (a \succ b),$$
where $x,y \in M$ and $a,b \in L$
turns $M \otimes L$ into an associative algebra.
\end{prop}

\subsection{Coding rooted planar binary trees}
\label{bintree}
Magmatic algebras consist of $K$-vector spaces equipped with one binary operation. They have been investigated by R. Holkamp \cite{H}, see also \cite{HLR}.
The operad is called $Mag$. It has been proved that the free magmatic algebra on one generator, $Mag(K)$ is constructed over rooted planar binary trees and the grafting, denoted by $\vee$, as operation.
By $Y_n$, we mean the set of rooted planar binary trees with $n$ nodes. Recall $\card(Y_n) =c_n$, the Catalan numbers. In low dimensions, these sets are:
$$Y_0:=\{ \treeO \}, \ Y_1:=\{ Y:=\treeA \}, \ Y_2:=\{ \treeAB, \treeBA \}, \ Y_3:=\{ \treeABC, \treeBAC, \treeACA, \treeCAB, \treeCBA \}.$$
For instance the grafting of $\treeA$ by itself
yields $\treeA \vee \treeA =\treeACA $. Therefore, as a $K$-vector space
$Mag(K)= \bigoplus_{n \geq 0} \ KY_n$.
Forgetting the operation $\succ$, the space $(K\mathcal{L}, \prec)$ can be seen as a magmatic algebra. Conversely,
the space $(Mag(K), \vee, 0)$ can be viewed as a L-algebra.
Consequently, if we denote the canonical injections $i: K \rightarrow Mag(K); \ 1_K \mapsto \treeO$ and $i': K \rightarrow K\mathcal{L}; \ 1_K \mapsto 1$, there exist a unique L-algebra morphism $\Phi$  and a unique magmatic morphism $\Psi$ such that the following diagrams commute:

$
\begin{array}{ccc}
&i & \\
K& \rightarrow & Mag(K) \\
  & & \\
& i' \searrow  & \downarrow \Psi \\
  & & \\
& & K\mathcal{L},
\end{array}
$
\quad \quad \quad  \quad \quad \quad \quad \quad  \quad \quad
$
\begin{array}{ccc}
&i' & \\
K& \rightarrow & K\mathcal{L} \\
  & & \\
& i \searrow  & \downarrow \Phi \\
  & & \\
& & Mag(K).
\end{array}
$

\noindent
As $\Phi$ is also a magmatic morphism, we get $\Phi \circ \Psi= id_{Mag(K)}$. Otherwise stated, rooted planar binary trees can be coded in a unique way by rooted symmetric planar ternary trees or simpler, by words made of 1 and $\bar{2}$. In low dimensions, we get:
$$ \Psi(\treeA)=\Psi(\treeO \vee \treeO)=\Psi(\treeO) \prec \Psi(\treeO)= 1 \prec 1= 111\bar{2}.$$
$$ \Psi(\treeAB)=\Psi(\treeA \vee \treeO)=\Psi(\treeA) \prec \Psi(\treeO)= 111\bar{2} \prec 1= 111\bar{2}11\bar{2}.$$
$$ \Psi(\treeBA)=\Psi(\treeO \vee \treeA)=\Psi(\treeO) \prec \Psi(\treeA)= 111\bar{2} \prec 1= 11111\bar{2}\bar{2}.$$
Here is a way to code straigthforwardly a binary trees into a word of $K\mathcal{L}$. Observe that in a word of $\mathcal{L}_{n+1}$, there are $2n+1$ times 1 and $n$ times $\bar{2}$ and in a binary tree of $Y_n$, there are $n+1$ leaves and $n$ nodes. Therefore, assign to each leaf or node of a binary tree $t$ a 1 and to each node a $\bar{2}$. Start with the node giving the most left leaf. This will give you (1)~1 (word) $\bar{2}$ (we put into parentheses the code of the left leaf and the right leaf). Now go a step below to meet another node. This will give you
((1) 1 (word) $\bar{2}$) 1 (word') $\bar{2}$ and so on, because $\Psi(t=t_{left} \vee t_{right}) = \Psi(t_{left})1 \Psi(t_{right})\bar{2}$.\\
For instance,

\begin{center}
\includegraphics*[width=8.5cm]{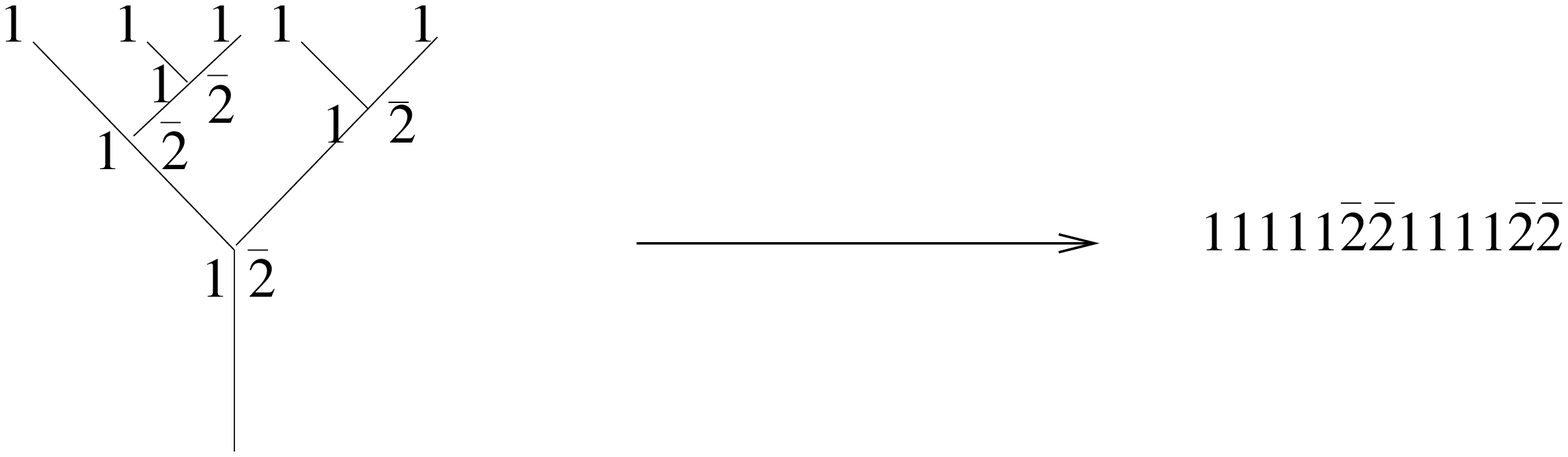}
\end{center}
\subsection{Free L-monoids and an arithmetics over trees}
\label{monoid}
Pursuing an idea of J.-L. Loday \cite{Lodayscd, Lodayarithm, Lerden}, we can propose an arithmetics from the $L$-operad\footnote{In \cite{Lerden}, such an arithmetics over the free associative $L$-algebra (now called duplicial-algebra) has been studied in relations with dendriform algebras.}. Indeed since the $L$-operad comes from a set operad, one can define over $\bigcup_{n>0} \mathcal{L}_n$ (disjoint unions) two nonassociative gradded additions and a multiplication as follows:
$$ l +_{_\succ} l' := 1l \bar{2} l', \ \ \ \ \ \
 l +_{_\prec} l' := l1l' \bar{2}, \ \ \ \ \ \
l \ltimes  l':= \varpi(l) \leftarrow l',$$
where $\varpi(l) \leftarrow l'$ means that the word 1 has to be replaced by $l'$ in $\varpi(l)$. For instance if $l:=1 \prec 1$,
then $l \ltimes  l'= l' \prec  l'$. As expected
$+_{_\succ}, \ +_{_\prec}: \mathcal{L}_n \times \mathcal{L}_m \rightarrow \mathcal{L}_{n+m}$ and
$\ltimes: \mathcal{L}_n \times \mathcal{L}_m \rightarrow \mathcal{L}_{nm}$, thus words from
$\mathcal{L}_{p}$, with $p$ a prime number, are prime for this arithmetics. Observe that $\ltimes$ is associative and left distributive with regards to additions.
L-monoids being straightforward to define, we get the following.
\begin{theo}
The free L-monoid over a set $X$ is given by,
$$ \mathcal{L}=(\bigcup_{n>0} \mathcal{L}_n \times (\underbrace{X \times \ldots \times X)}_{n \  \textrm{copies}}, +_{_\succ}, \ +_{_\prec}).$$
Moreover, for any words $l,l',l''$,
$$ l +_{_\bullet} l' = l +_{_\bullet} l'' \Leftrightarrow l'=l''; \ \ \ \ \ l' +_{_\bullet} l = l'' +_{_\bullet} l \Leftrightarrow l'=l'', \ \ \textrm{where}  \ \bullet \ = \ \prec, \ \succ.$$
$$ l' \ltimes  l = l'' \ltimes l \Leftrightarrow l'=l''.$$
\end{theo}
\Proof
The first two assertions are a straightforward consequence of the structure of free L-algebra. As the L-algebra $K\mathcal{L}$ is free, there exists a unique L-algebra automorphism sending the generator 1 to $l$, therefore since the multiplication is left distributive, the map $- \ltimes l: K\mathcal{L} \rightarrow  K\mathcal{L}$ is this automorphism, hence the third assertion.
\eproof

\subsection{Associative algebras from L-algebras}
\label{as}

Observe that associative algebras are  L-algebras hence a functor $inc:\textsc{As} \rightarrow \textsc{L-alg}$.
Let $(L, \prec, \succ)$ be a L-algebra and $I$ be the L-ideal generated by the relations $x\succ y- x\prec y$ for $x,y \in L$. Then $L/I$ is an associative algebra, hence a functor $F_{As}: \textbf{L-alg} \rightarrow \textbf{As}$. Denote by $\pi:L \twoheadrightarrow L/I$ the canonical surjection. Let $A$ be an associative algebra. For any morphism of L-algebras $f:L \rightarrow A$ there exists a unique morphism of associative algebras $\tilde{f}: L/I \rightarrow A$ such that $f=\tilde{f} \circ \pi$. Similarly,
if $\tilde{f}: L/I \rightarrow A$ is a morphism of associative algebras then $f:=\tilde{f} \circ \pi:L \rightarrow inc(A)$ is a morphism of L-algebras. Hence,
$$ Hom_{As}(F_{As}(L);A) \cong Hom_{L-alg}(L; inc(A)),$$
the functor $F_{As}$ is left adjoint to the functor $inc$.
\subsection{Homology of L-algebras}
\label{homology}
By L$^!(n)$, we mean the $K$-vector space spanned by $n$-ary operations of the L$^!$-operad made out of our two generating binary operations $\vdash$ and $\dashv$. We proved that L$^!(n)=0$ as soon as $n>3$. We have L$^!(1)=K.id$, L$^!(2)=K.\vdash \oplus K.\dashv$ and L$^!(3)=K.( \vdash) \dashv$. Following V. Ginzburg and M. Kapranov, the chain-complex over a L-algebra $A$ is   restricted to
$$ L^!(3)\otimes A^{\otimes 3} \xrightarrow{d} L^!(2)\otimes A^{\otimes 2} \xrightarrow{d}
L^!(1)\otimes A,$$
where $d$ is the differential operator which agrees in low dimensions with the L-algebra structure of $A$.

Therefore, so as to give explicitly an homology theory for L-algebras let us define
$C_3$ to be the set $\{1\}$; $C_2$ to be the set $\{1,2\}$; $C_1$ to be another copy of the set $\{1\}$. Let $A$ be a L-algebra. The module of $n$-chains, for $n=1,2,3$, is $C^L_n(A):=KC_n \otimes A^{\otimes n}$. The differential operator $d$ is defined as follows:
$$ d(1; x \otimes y \otimes z):=(1; (x\succ y) \otimes z)- (2;x \otimes (y \prec z));$$
$$d(1; x \otimes y):=(1; x\prec y);$$
$$d(2; x \otimes y):=(1; x\succ y),$$
for all $x,y,z \in A$.
We do have $d^2=0$ since,
$$d^2(1; x \otimes y \otimes z):=d(1; (x\succ y) \otimes z)- d(2;x \otimes (y \prec z))=(1;( x\succ y)\prec z)- (1; x \succ (y \prec z))=0.$$
Hence the complex,
$$ CL_*(A): \ \ \ 0 \xrightarrow{d} KC_3 \otimes A^{\otimes 3}\xrightarrow{d} KC_2 \otimes A^{\otimes 2} \xrightarrow{d} KC_1 \otimes A\xrightarrow{d} 0.$$
By definition, the homology of the L-algebra $A$ is the homology of our short chain-complex $CL_*(A)$ and,
$$ HL_n(A):=H_n(CL_*(A),d); \ n=1,2,3.$$
We get $ HL_1(A):=A/J$, where $J$ is the ideal generated by the $x\prec y$ and $x \succ y$, for $x,y \in A$.

\begin{theo}
Let $V$ be a $K$-vector space and L$(V)$ be the free L-algebra over $V$. Then,
$$ HL_1(L(V)) \simeq V,$$
$$ HL_n (L(V))=0, \ \ \textrm{for} \ n>1.  $$
Therefore, the L-operad is Koszul.
\end{theo}
\Proof
As the L-operad is regular, we restrict the proof to the free L-algebra on one generator, \textit{i.e.}, $K\mathcal{L}$.
To ease notation, we rename $d$ as follows:
$$ CL_*(K\mathcal{L}): \ \ \ 0 \xrightarrow{d_3} KC_3 \otimes K\mathcal{L}^{\otimes 3}\xrightarrow{d_2} KC_2 \otimes K\mathcal{L}^{\otimes 2} \xrightarrow{d_1} KC_1 \otimes K\mathcal{L}\xrightarrow{d_0} 0.$$
The first assertion is trivial since $\Im d_1 = \bigoplus_{n \geq 2} \ K\mathcal{L}_n$ and $\ker d_0= K\mathcal{L}$, thus $\ker d_0/ \Im d_1\simeq K$. Let us show that $HL_3 (K\mathcal{L})=0,$ that is $d_2$ is injective. Let $l,l',l''$ be three words such that $d_2(1; l \otimes l' \otimes l'')=0$. Hence, we get both $(1; (l\succ l') \otimes l'')=0$ and $(2; l  \otimes (l'\prec l''))=0$. As operations respect the graduation, one of them has to be equal to zero. Let us show that $HL_2 (K\mathcal{L})=0,$ that is $\Im d_1=\ker d_2$. We know that $\ker d_1 \supset \Im d_2$. Let $l_1,l_2,l_3,l_4$ be four words
and $\lambda, \mu \in K$ such that $d_1(\lambda (1; l_1 \otimes l_2) -\mu (2; l_3 \otimes l_4))=0$. We get $\lambda (1; l_1 \prec l_2) = \mu (1; l_3 \succ l_4)$.
Hence, $\mu = \lambda$ and $l_1 \prec l_2=l_3 \succ l_4$, that is:
$$ w:= l_1 1 l_2 \bar{2}= 1l_3 \bar{2} l_4.$$
The proof of Theorem~\ref{freeee} shows that such an equality holds if and only if there exists a word $u$ such that
$l_1= l_3 \succ u.$
Consequently, the word $l_4$ can be written as:
$l_4= u \prec l_2.$
Therefore, the expression we started with,
$\lambda (1; l_1 \otimes l_2) -\mu (2; l_3 \otimes l_4) \in \ker d_1$ can be written
$ \lambda d_2(1; l_3 \otimes u \otimes l_2)$.
Our complex is exact and the L-operad is Koszul.
\eproof

\noindent
As the L-operad is Koszul, applying results of \cite{GK} gives the following.
\begin{enumerate}
 \item{The L$^!$-operad is Koszul.}
\item{ The generating functions of the L$^!$-operad and the L-operad are inverse one another for the composition, that is:
$$ f_L(-f_{L^!}(-x))=x.$$
This gives an algebraic interpretation of the fact that these two series have been discovered to be inverse one another in $A006013$ from the On-Line Encyclopedia of Integer Sequences. }
\end{enumerate}
\section{On an action of the unit}
\label{action}
Recall $1_K$ denotes the unit of the field $K$. Let $(L, \prec, \succ)$ be a  L-algebra. Recall \cite{Coa}, directed graphs equipped with probability vectors, viewed as (Markov) L-coalgebras naturally have a left counit equals to a right one. Dually, left and right units can be introduced. We focus on the case when they coincide.
Over $K \oplus L$ a structure of $L$-algebra can be constructed as follows.
$$ 1_K\prec 1_K=1_K = 1_K\succ 1_K;$$
$$ \forall x \in L; \ \ 1_K \prec x=t(x); \ \ x \succ 1_K=s(x); \ \ \ 1_K\succ x=x= x\prec 1_K,$$
where $s,t:L \rightarrow L$ are linear maps such that,
$$ \forall x,y \in L; \ \ \ t(x) \prec y= x \succ s(y).$$
A L-algebra is said to be unital if it has an element denoted by $1$ and a pair $(t,s)$ verifying the above equations. If $L$ and $L' $ are (unital) L-algebras then the following operations,
$$ (x \otimes x') \prec (y \otimes y')= (x\prec y) \otimes (x' \prec y');$$
$$ (x \otimes x')\succ (y \otimes y')= (x \succ y) \otimes (x' \succ y'), $$
for $x,y \in L$ and $x',y' \in L'$,
turns $L \otimes L'$ into a (unital) L-algebra.
\begin{prop}
Let $V$ be a $K$-vector space. There exits a cocommutative coassociative coproduct and a counit over
the augmented free L-algebra  $K \oplus \textrm{L}(V)$ which are unital L-algebra morphisms.
\end{prop}
\Proof
Fix $l_0 \in \textrm{L}(V)$.
Define the map $s,t: \textrm{L}(V) \rightarrow \textrm{L}(V)$ by,
$$ s(l):=l_0 \prec l, \ \ \textrm{and} \ \ \ t(l):= l \succ l_0.$$
Following J.-L. Loday \cite{Lodayscd}, since $\textrm{L}(V)$ is free, the map
$v \mapsto 1_K \otimes v + v \otimes 1_K$, for any $v \in V$, has a natural extention, morphism of L-algebras, $\Delta:K \oplus \textrm{L}(V) \rightarrow (K \oplus \textrm{L}(V))^{\otimes 2}$ which is coassociative. As the flip operator is a L-morphism and leaves $v \mapsto 1_K \otimes v + v \otimes 1_K$ invariant, the coproduct $\Delta$ will be cocommutative.
The counit $\epsilon$ is such that $\epsilon(\textrm{L}(V))=0$.
\eproof

\NB
For instance,
$\delta(l \prec l')= 1_K \otimes (l \prec l') + (l \prec l') \otimes 1_K + l \otimes t(l') + t(l) \otimes l'$.
Equipped with this coproduct, $K \oplus K\mathcal{L}(V)$ is not connected in the sense of Quillen except if $s=t=0$.
In this case, the $K$-vector space of the primitive elements $Prim(K\mathcal{L}(V))$ is a L-algebra isomorphic to $K\mathcal{L}(V)$.

\section{Structure theorems and nonunital infinitesimal relations}
\label{structure}
The aim of this section is to obtain a general structure theorem which will be useful in Section~5.

\subsection{The structure theorem for coassociative $[n]-Mag$-bialgebras}
\label{mag}
We first start with recalling what the nonunital infinitesimal compatibility relation over a binary  operad $\mathcal{P}$ is, allowing the definition of coassociative $\mathcal{P}$-bialgebras.
We generalize the magmatic operad $Mag$ to $[n]-Mag$ the magmatic operad with $n$ operations and some results obtained in \cite{HLR}. Inspired by proofs from \cite{HLR}, we show the existence of a structure theorem for the triple $(As, [n]-Mag, Prim \ [n]-Mag)$ and obtain, as a consequence, a structure theorem for the triple $(As, \textrm{L}, Prim \ \textrm{L})$. Let us start with the definition of coassociative $\mathcal{P}$-bialgebras.

\begin{defi}{}
Let $\mathcal{P}$ be a binary, quadratic operad.
A coassociative $\mathcal{P}$-bialgebras, $As^c-\mathcal{P}$-bialgebra for short, is a $\mathcal{P}$-algebra $P$ equipped with a coassociative coproduct $\delta:P \rightarrow P^{\otimes 2}$ verifying the so-called nonunital infinitesimal compatibility relation,
$$(*) \ \ \ \ \ \ \ \ \delta(x \bullet y)= \delta(x)\bullet y + x \bullet \delta(y) + x\otimes y, \ \ i.e.,$$
$$(*) \ \ \ \ \ \ \ \ \delta(x \bullet y)= x_{(1)} \otimes (x_{(2)}\bullet y) +  (x \bullet y_{1}) \otimes y_{2} + x \otimes y,$$
where $\delta(x)= x_{(1)} \otimes x_{(2)}$ (Sweedler's notation), for any $x,y \in P$ and any generating operations $\bullet  \in \mathcal{P}(2)$.
An $As^c-\mathcal{P}$-bialgebra $\mathcal{H}$ is said to be connected if $\mathcal{H}=\bigcup_{r \geq 0} \ F_r \mathcal{H},$ where $F_r \mathcal{H}$ is the coradical filtration of $\mathcal{H}$ defined recursively by,
$$F_0 \mathcal{H}:=K.1_K, \ \ F_r \mathcal{H}:=\{x \in \mathcal{H} \ \vert \  \delta(x) \in  F_{r-1} \mathcal{H} \otimes F_{r-1} \mathcal{H} \}.$$
By definition the space of primitive elements is defined as,
$$ Prim \ \mathcal{H}:= \ker \delta.$$
\end{defi}

\noindent
\textbf{On the cofree coalgebra.}
Let $V$ be a $K$-vector space. Recall that $As^c(V):= \bigoplus_{m>0} \ V^{\otimes m}$ as a $K$-vector space and  equipped with the deconcatenation coproduct $\hat{\delta}$, defined by,
$$ \hat{\delta}(v):=0, \ \ \ \ \forall \ m>1, \ \ \hat{\delta}(v_1 \otimes \ldots \otimes v_m):= \sum_{i=1}^{m-1} \ (v_1 \otimes \ldots \otimes v_i) \otimes (v_{i+1} \otimes \ldots \otimes v_m),$$
is the cofree connected coassociative coalgebra in the corresponding category. Recall also that $\hat{\delta}$ verifies Formula $(*)$.

\noindent
\textbf{The magmatic operad with $n$ operations.} Fix an integer $n>0$.
Let $[n]-Mag$ be the (free) operad generated by $n$ binary operations $\bullet_i, \ i=1, \ldots n$, that is $[n]-Mag(2)=K\{ \bullet_i, \ i=1, \ldots n\}$. We set $[1]-Mag:=Mag$.
Its dual is the operad $[n]-Nil$ also generated by $n$ binary operations $\bullet^{\vee}_i, \ i=1, \ldots, n$ and such that any nontrivial compositions vanish, i.e., $[n]-Nil(m)=0$ for $m>2$.
Therefore, $[n]-Mag$ is a Koszul operad.
For $p\geq 0$, let $Col[n]Y_p$ be the set of rooted planar binary trees whose nodes are colored by a color $i=1, \ldots, n$.
For instance, $Col[n]Y_0=\{ \treeO \}$, $Col[n]Y_1=\{ \treeA_i, \ i=1 \ldots n \}$.
The $n$-grafting operations we are looking for are denoted by $\vee_i, \ i=1, \ldots, n$. Hence $t \vee_i t'$ means that the tree $t$ is grafted to $t'$ \textit{via} a root colored by $i$.
The free $[n]-Mag$-algebra over a $K$-vector space $V$ is the $K$-vector space,
$$ [n]-Mag(V)=\bigoplus_{p\geq 0} KCol[n]Y_{p} \otimes V^{\otimes (p+1)},$$
equipped with the $n$-operations,
$$ (t; v_1 \otimes \ldots \otimes v_{p+1}) \vee_i (s; w_1 \otimes \ldots \otimes v_{q+1})= (t \vee_i s; v_1 \otimes \ldots \otimes v_{p+1}\otimes w_1 \otimes \ldots \otimes v_{q+1}).$$
\begin{prop}
Let $[n]-Mag(V)$ be the free $[n]-Mag$-algebra over the $K$-vector space $V$. Then, there exists a unique coproduct $\delta$ vanishing on $K\treeO \otimes V \simeq V$ and turning $[n]-Mag(V)$ into an $As^c-[n]-Mag$-bialgebra.
\end{prop}
\Proof
Equip $[n]-Mag(V) \otimes [n]-Mag(V)$ with the following $n$ operations,
$$ (x \otimes y) \vee_i (x' \otimes y')=
(x \vee_i x') \otimes (y \vee_i  y'), $$
hence $[n]-Mag(V) \otimes [n]-Mag(V)$ is also a $[n]-Mag$-algebra.
Define now the coproduct $\delta:[n]-Mag(V) \rightarrow  [n]-Mag(V) \otimes [n]-Mag(V)$ as follows.
$$ \delta(\treeO; v)=0,$$
for any $v\in V$ and recursively by the Formula $(*)$. For instance,
$\delta((\treeA_i; v \otimes w))=\delta((\treeO; v) \vee_i (\treeO; w))= v \otimes w$ and so on. It is straightforward to prove recursively that Formula $(*)$ implies the  coassociativity of $\delta$. Hence, the uniqueness of $\delta$ and
$[n]-Mag(V)$ has $As^c-[n]-Mag$-bialgebra structure.
\eproof

\noindent
To show that the $As^c-[n]-Mag$-bialgebra is connected, we use the following lemma adapted from \cite{HLR}. First of all, let $t$ be a colored planar binary rooted tree with $p+1$ leaves, numbered from left to right by $1,2, \ldots, p+1$. Let $j=1, \ldots,p$. We split the tree $t$ into two colored trees $t^j_{(1)}$ and $t^j_{(2)}$ as follows. The tree $t^j_{(1)}$ is the part of $t$ at the left hand side of the path from the leaf $j$ to the root, the path being included. The tree $t^j_{(2)}$ is the part of $t$ at the right hand side of the path going from the leaf $j+1$ to the root. The color assigned to the node ``father'' of the leaves $j$ and $j+1$ is then ignored.
For instance,
\begin{center}
\includegraphics*[width=7.5cm]{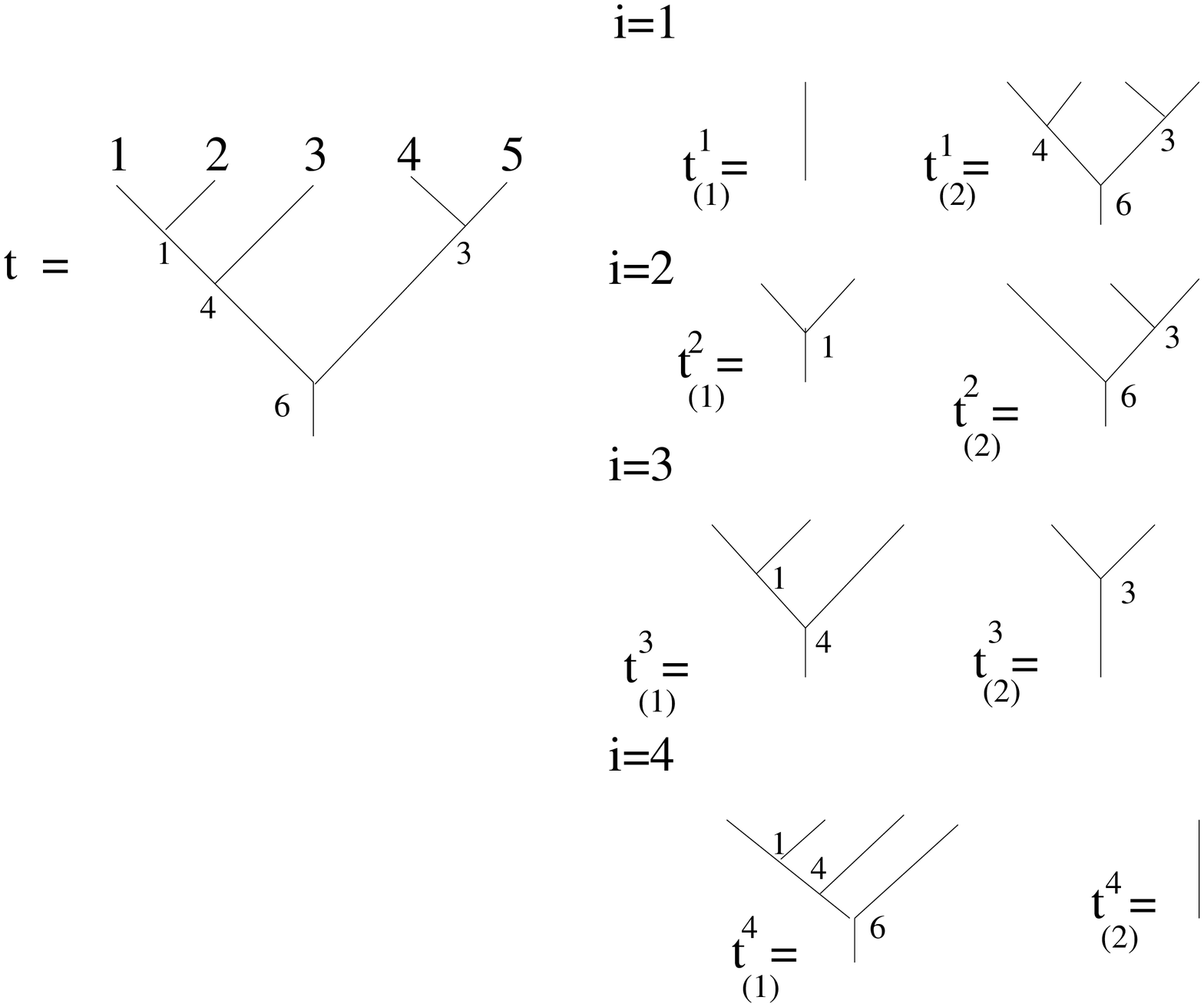}
\end{center}
\begin{prop}
Let $[n]-Mag(V)$ be the free $[n]-Mag$-algebra over the $K$-vector space $V$. Then, the coproduct $\delta$ can be written as a sum of co-operations $\sum_{1 \leq j \leq p} \ \delta_j$, where,
$$ \delta_j(t; v_1 \otimes \ldots \otimes v_{p+1})= (t^j_{(1)}; v_1 \otimes \ldots \otimes v_{j})\otimes (t^j_{(2)}; v_{j+1} \otimes \ldots \otimes v_{p+1}).$$
Hence, $([n]-Mag(V), \ \delta)$ is a connected $As^c-[n]-Mag$-bialgebra.
\end{prop}
\Proof
We adapt the proof from \cite{HLR}.
Define the co-operation $\Delta$ by $\Delta:=\sum_{1 \leq j \leq p} \ \delta_j$, where,
$$ \delta_j(t; v_1 \otimes \ldots \otimes v_{p+1})= (t^j_{(1)}; v_1 \otimes \ldots \otimes v_{j})\otimes (t^j_{(2)}; v_{j+1} \otimes \ldots \otimes v_{p+1}).$$
The co-operation $\Delta$ vanishes on $\treeO \otimes v$, for all $v \in V$.
Let us show that $\Delta$ verifies Formula $(*)$.
For $p>2$, there exist a unique $k=1, \ldots p-1$, a unique $i=1, \ldots, n$, unique colored rooted planar binary trees $t^{l} \in Col[n]Y_{k}$ and $t^{r}\in Col[n]Y_{p-k-1}$ such that $t=t^{l} \vee_i t^{r}$.
Hence,
$$ \Delta((t^{l}; v_1 \otimes \ldots \otimes v_{k+1}) \vee_i (t^{r}; v_{k+2} \otimes \ldots \otimes v_{p+1}))=\Delta(t; v_1 \otimes \ldots \otimes v_{p+1})=$$
$$ \sum_{j=1}^{k-1} ((t^l)^j_{(1)}; v_1 \otimes \ldots \otimes v_{j})\otimes ((t^l)^j_{(2)}\vee_i \ t^r; v_{j+1} \otimes \ldots \otimes v_{p+1})+ (t^{l}; v_1 \otimes \ldots \otimes v_{k+1}) \otimes (t^{r}; v_{k+2} \otimes \ldots \otimes v_{p+1})$$
$$+ \sum_{j=k+2}^{p} (t^l \vee_i (t^r)^{j-k-1}_{(1)}; v_1 \otimes \ldots \otimes v_{j})\otimes ((t^r)^{j-k-1}_{(2)}; v_{j+1} \otimes \ldots \otimes v_{p+1})=$$
$$ \Delta((t^{l}; v_1 \otimes \ldots \otimes v_{k+1})) \vee_i (t^{r}; v_{k+2} \otimes \ldots \otimes v_{p+1}) + (t^{l}; v_1 \otimes \ldots \otimes v_{k+1}) \vee_i \Delta(t^{r}; v_{k+2} \otimes \ldots \otimes v_{p+1})$$
$$+ (t^{l}; v_{k+2} \otimes \ldots \otimes v_{p+1}) \otimes  (t^{r}; v_{k+2} \otimes \ldots \otimes v_{p+1}).$$
Therefore, $\Delta$ verifies the formula $(*)$.
Consequently, the previous proposition shows that $\Delta$ is coassociative and thus has to coincide on the whole $[n]-Mag(V)$ with $\delta$. Moreover, this formula shows that $[n]-Mag(V)$ is connected.
\eproof

We now refer to notation, definitions and results of J.-L. Loday \cite{GB}. Since the nonunital infinitesimal relation $(*)$ is distributive (Hypothesis $H0$ \cite{GB}) and since $[n]-Mag(V)$ is equipped with a coassociative  $[n]-Mag(V)$-bialgebra (Hypothesis $H1$ \cite{GB}),
$Prim \ [n]-Mag$ is also an operad, suboperad of $[n]-Mag$. Therefore, it makes sense to deal with $Prim \ [n]-Mag$-algebras and the forgetful functor,
$$ F: \ [n]-Mag \mapsto Prim \ [n]-Mag,$$
has a left adjoint, the so-called universal enveloping algebra functor,
$$ U: \ Prim \ [n]-Mag \mapsto [n]-Mag.$$
\begin{theo}
For any $As^c-[n]-Mag$-bialgebra, $\mathcal{H}$, the following are equivalent:
\begin{enumerate}
\item {The $As^c-[n]-Mag$-bialgebra $\mathcal{H}$ is connected.}
\item {There is an isomorphism of bialgebras $\mathcal{H} \simeq U(Prim \mathcal{H})$.}
\item{There is an isomorphism of connected coalgebras $\mathcal{H} \simeq As^c(Prim \mathcal{H})$,}
\end{enumerate}
That is, the triple of operads $(As, [n]-Mag, Prim([n]-Mag))$ is good.
\end{theo}
\Proof
We will apply Theorem 2.5.1. from \cite{GB} by checking that the so-called Hypothesis $(H2epi)$ \cite{GB} holds.
Let $V$ be a $K$-vector space. By the previous proposition, $[n]-Mag(V)$ has a natural $As^c-[n]-Mag$-bialgebra structure.  Consequently, the projection map $proj_V:
[n]-Mag(V) \twoheadrightarrow V$ and the cofreness of $As^c(V)$ give a unique coalgebra map:
$$ \phi(V): [n]-Mag(V) \rightarrow As^c(V),$$
such that $\pi \circ \phi(V) = proj_V,$
where $\pi: As^c(V) \twoheadrightarrow V$ is the canonical projection.
The formula giving the coproduct $\delta$ implies that $\phi(V)(t)=1_K$ for any colored tree $t$ (recall $As^c_n=K$). Thus, $\phi(V)$ is surjective.
Fix now a color $i$ and denote by $comb_m$, $m \geq 0$, the left comb obtained recursively by $comb_0:= \treeO$ and $comb_m:=comb_{m-1} \vee_i \treeO.$
Define the map $s(V): As^c(V) \rightarrow
[n]-Mag(V)$ by $1_K \in As^c_m \mapsto comb_{m-1} \in [n]-Mag_{m}$. Then, $s(V)$ is a coalgebra map since,
$$ \delta(comb_0; v)=0,$$
$$ \forall \ m>1, \ \ \delta(comb_m; v_1 \otimes \ldots \otimes v_{m+1}) = \sum_{j=1}^{m-1} \ (comb_{j-1}; v_1 \otimes \ldots \otimes v_{j}) \otimes (comb_{m-j}; v_{j+1} \otimes \ldots \otimes v_{m+1}),$$
and $\phi(V) \circ s(V)= id_{As^c(V)}$.
Hence, the natural coalgebra map $\phi(V)$ is surjective and admits a natural coalgebra splitting $s(V)$, hence Hypothesis $(H2epi)$ holds.
Therefore, applying \cite{GB} Theorem 2.5.1., the triple $(As, [n]-Mag, Prim([n]-Mag))$ is a good triple of operads.
\eproof

\begin{theo}
\label{outil}
Let $\mathcal{P}$ be a binary, quadratic operad such that for any generating operations $\bullet_i \in \mathcal{P}(2)$, there exist relations only in $\mathcal{P}(3)$ of the form,
$$ (++) \ \ \ \sum_{i,j; \ \sigma_{i,j}\in S_3} \ \lambda_{i,j} \ \bullet_j( \bullet_i \otimes id) \sigma_{i,j} = \sum_{i,j; \ \sigma_{i,j}\in S_3} \ \lambda_{ij} \ \bullet_i  (id \otimes \bullet_j )\sigma_{i,j} ,$$
for any $i,j \in \{1, \ldots, \dim \mathcal{P}(2) \}$ and $\lambda_{ij} \in K$. Then, the triple $(As, \mathcal{P}, Prim \mathcal{P})$ is good. Considering only binary quadratic operad coming from a set operad, quadratic relations of the form:
$$(+++) \ \ \ \bullet_j( \bullet_i \otimes id)\sigma_{i,j}=(id \otimes \bullet_j )\sigma_{i,j},$$
are the only ones giving such good triples.
\end{theo}
\Proof
Suppose first $\mathcal{P}$ to be regular.
Set $n=\dim \mathcal{P}_2$. In the operad $[n]-Mag$, the operations $\bullet_j  (\bullet_i \otimes id)  - \bullet_i  (id \otimes \bullet_j) \in [n]-Mag_3$ are primitive operations.
Indeed,
let $x,y,z$ be primitive elements of a $As^c-[n]-Mag$-bialgebra $(\mathcal{H}, \ \delta)$. Then,
$$ \delta((x\bullet_i y) \bullet_j z)= x \otimes (y \bullet_j z) + (x\bullet_i y) \otimes z, $$
which is compensated with,
$$ \delta(x\bullet_i (y \bullet_j z))= (x\bullet_i y) \otimes z+ x \otimes (y \bullet_j z).  $$
Therefore,
$(x\bullet_i y) \bullet_j z - x\bullet_i (y \bullet_j z) \in \ker \delta$ and any linear combinations of such relations remains primitive operations. This result still stands if one permutates the entries in the same way on each side of Formula $(++)$. The relations $(+++)$ are the only ones to give primitive operations when only operads coming from set operads are considered since to characterize $\mathcal{P}$, linear combinations of ternary operations are not allowed.
Denote by $J$ the operadic ideal generated by the primitive operations,
$$\sum_{i,j; \ \sigma_{i,j}\in S_3} \ \lambda_{i,j} \ \bullet_j( \bullet_i \otimes id) \sigma_{i,j} - \sum_{i,j; \ \sigma_{i,j}\in S_3} \ \lambda_{ij} \ \bullet_i  (id \otimes \bullet_j )\sigma_{i,j}.$$
Applying Proposition 3.1.1. of J.-L. Loday \cite{GB}, one gets that $(As^c,[n]-Mag/J, Prim [n]-Mag/J )$ is still a good triple of operads.
As, $[n]-Mag/J\simeq \mathcal{P}$, the triple of operads,
$(As, \mathcal{P}, Prim \mathcal{P})$ is good.
\eproof

\begin{exam}{}
The structure theorem holds for the $As^c$-duplicial-bialgebras as proved in a different way in \cite{GB}. As another example, in an associative algebra, consider two associative products $\star_1$ and $\star_2$. Requiring the product $\star = \star_1 + \star_2$ to be associative leads to the so-called Hochschild 2-cocycle \cite{Gerst1}:
$$(**) \ \  \ \ \ \   (x \star_1 y) \star_2 z + (x \star_2 y) \star_1 z= x \star_1 (y \star_2 z) + x \star_2 (y \star_1 z). $$
Consider the regular, binary, quadratic operad $G$
made out with two associative products verifying the relation $(**)$, then Theorem \ref{outil} claims the existence of a notion of $As^c-G$-bialgebras and a good triple $(As, G, Prim G)$. The triple $(As,Pre-Lie, Prim Pre-Lie)$ is another (well-known) example. The relations $(++)$ are not the only ones to give such good triples (if the operad $\mathcal{P}$ does not come from a set operad). For instance the quadratic binary operad defined to be $[4]-Mag$ (4 binary operations $\bullet_1, \bullet_2,\bullet_3,\bullet_4$)
divided out by the operadic ideal generated by, $$(\bullet_3-\bullet_4)((\bullet_1- \bullet_2) \otimes id),$$
will also provide a good triple.
\end{exam}

\subsection{Idempotents and primitive elements }
\label{idemp}
We now improve a result of \cite{LodRon}.
Let $\mathcal{P}$ be a binary operad for which the notion of connected $As^c-\mathcal{P}$-bialgebras stands. Let $(\mathcal{H}, \delta)$ be such a bialgebra. For each generating operation $\bullet \in \mathcal{P}(2)$, define the linear map, ${e^r}_{\bullet}: \mathcal{H} \rightarrow \mathcal{H},$
recursively by,
$$x \mapsto {e^r}_{\bullet}(x):= x - x_{(1)} \bullet {e^r}_{\bullet}(x_{(2)}),$$
where using Sweedler's notation $\delta(x):= x_{(1)}\otimes x_{(2)}$.
As $\mathcal{H}$ is connected, the map ${e^r}_{\bullet}$ is well defined. Similarly, define the linear map, ${e^l}_{\bullet}: \mathcal{H} \rightarrow \mathcal{H},$
recursively by,
$$x \mapsto {e^l}_{\bullet}(x):= x - {e^l}_{\bullet}(x_{(1)}) \bullet x_{(2)}.$$
\begin{prop}
For each generating operation $\bullet \in \mathcal{P}(2)$,
the maps ${e^r}_{\bullet}$ and ${e^l}_{\bullet}$ are idempotents from $\mathcal{H}$ to $Prim \mathcal{H}$.
Suppose the existence of an associative generating operation $\star \in \mathcal{P}(2)$.
Then,
$$\ker e_{\star}=\mathcal{H} \star \mathcal{H}.$$
\end{prop}
\Proof
We focus on ${e^r}_{\bullet}$ renamed in $e_{\bullet}$.
We will prove these claims by induction on the filtration $F_n \mathcal{H}$ of $ \mathcal{H}$.
Let $x \in Prim \mathcal{H}$, then $\delta(x)=0$, thus $e(x)=x$ and $e(x) \in Prim \mathcal{H}$. Let $x \in F_n \mathcal{H}$ and suppose $e_{\bullet}(y) \in Prim \mathcal{H}$ for any $y \in F_r  \mathcal{H}$, $r< n$. As $ \mathcal{H}$ is connected,
\begin{eqnarray*}
\delta(e_{\bullet}(x)) &=& \delta(x) - \delta(x_{(1)} \bullet e_{\bullet}(x_{(2)})) = \delta(x) - \delta(x_{(1)}) \bullet e_{\bullet}(x_{(2)}) -  x_{(1)} \otimes e_{\bullet}(x_{(2)}),\\   &=& \delta(x) - x_{(11)} \otimes (x_{(12)} \bullet e_{\bullet}(x_{(2)})) -  x_{(1)} \otimes e_{\bullet}(x_{(2)}).
\end{eqnarray*}
As $\delta$ is coassociative, this is equal to:
$$ \delta(x) - x_{(1)} \otimes (x_{(21)} \bullet e_{\bullet}(x_{(22)})) -  x_{(1)} \otimes e_{\bullet}(x_{(2)})= \delta(x) - x_{(1)} \otimes (id-e_{\bullet})(x_{(2)}) -  x_{(1)} \otimes e_{\bullet}(x_{(2)})=0.$$
Let $x \in \mathcal{H}$, then $e_{\bullet}(x)$ is primitive, therefore $e_{\bullet}(e_{\bullet}(x))=e_{\bullet}(x)$ and $e_{\bullet}$ is an idempotent.
Suppose the existence of an associative generating operation $\star \in \mathcal{P}(2)$.
Let $x,y \in \mathcal{H}$. To prove that
$e_{\star}(x \star y) =0$, we proceed by induction on the sum of the filtration-degrees of $x$ and $y$. If $x,y \in Prim \mathcal{H}$, then $\delta(x \star y)= x \otimes y$.
Therefore,
$e_{\star}(x \star y)= x \star y - x \star e_{\star}(y)= x \star y - x \star y=0$ since $e_{\star}$ is an idempotent. Suppose this result holds when the sum of the filtration-degrees is strictly less than the one of $x$ and $y$.
As $\delta(x \star y):= x \otimes y + x_{(1)} \otimes (x_{(2)} \star y)+ (x \star y_{(1)}) \otimes y_{(2)},$ and $\star$ is associative,
$$
e_{\star}(x \star y)= x \star y- [x \star e_{\star}(y) + (x \star y_{(1)})\star e_{\star} (y_{(2)})]= x \star y- x \star (e_{\star}(y)- y_{(1)} e_{\star}\star (y_{(2)}))= x \star y- x \star y=0,
$$
holds. Suppose $x \in \ker e_{\star}$, then $e_{\star}(x)=0$ and $x= x_{(1)}\star e_{\star}(x_{(2)})$. Therefore, $x \in \mathcal{H} \star \mathcal{H}$.
\eproof

\noindent
For each $n>0$, denote by $e_{\bullet;n}$ the restriction of $e_{\bullet}$ to $\mathcal{P}(n) \otimes_{S_n} V^{\otimes n}$.
\begin{prop}
\label{isomm}
Let $V$ be a $K$-vector space and a generating operation $\bullet \in \mathcal{P}(2)$. Then, for all $n$,
$$ \mathcal{P}(n) \otimes_{S_n} V^{\otimes n}/\ker e_{\bullet;n} \simeq (Prim \mathcal{P})(n)\otimes_{S_n} V^{\otimes n},$$
as $K$-vector spaces.
\end{prop}
\Proof
We show that,
$$e_{\bullet;n}: \mathcal{P}(n) \otimes_{S_n} V^{\otimes n} \rightarrow (Prim \mathcal{P})(n)\otimes_{S_n} V^{\otimes n},$$
is well-defined and surjective.
We proceed by induction on the degree of the involved $K$-vector spaces.
We have $e_{\bullet;1}(V)=V$. Let $\perp \in \mathcal{P}(2)$ be another generating operation. For $n=2$, if $x,y \in V$, then $e_{\bullet;2}(x \perp y)=  x \perp y - x \bullet y \in (Prim \mathcal{P})(2)$. Suppose the result holds up to a $n-1$. Then, for a monomial of $\mathcal{P}(n) \otimes_{S_n} V^{\otimes n}$, as $\mathcal{P}$ is binary, there exits two monomials of smaller degrees $X$ and $Y$ and a generating operation $\diamond \in \mathcal{P}(2)$ such that it can be written $X \diamond Y$. Therefore,
$$ e_{\bullet}(X \diamond Y):= X \diamond Y - X \bullet e_{\bullet}(Y),$$
hence maps $e_{\bullet;n}$ are well-defined by induction. To prove the restriction of the idempotent $e_{\bullet}$ on the homogeneous coponents is surjective, let $x \in (Prim \mathcal{P})(n)\otimes_{S_n} V^{\otimes n}$. As $e_{\bullet}$ is surjective, there exists a $y \in \mathcal{P}(V)$ such that $
e_{\bullet}(y)=x$. Therefore $y$ and $x$ have the same degree and $e_{\bullet;n}: \mathcal{P}(n) \otimes_{S_n} V^{\otimes n} \twoheadrightarrow (Prim \mathcal{P})(n)\otimes_{S_n} V^{\otimes n},$
factors through $\ker e_{\bullet}$.
\eproof

\NB
This result can be of assistance when searching a presentation of the operad $Prim \mathcal{P}$.

Let $V$ be a $K$-vector space and $\mathcal{P}(V)$ be the free $\mathcal{P}$-algebra over $V$. Suppose the existence of an involution $\dagger$ over $V$. The $\mathcal{P}$-algebra $\mathcal{P}(V)$ is said to be involutive if there exists a map still denoted by $\dagger$,
$$\dagger: \mathcal{P}(2) \rightarrow \mathcal{P}(2), \ \ \bullet_i \mapsto {\bullet_i}^\dagger,$$
given by induction by,
$$(t \bullet_i s)^\dagger= s^{\dagger} \ {\bullet_i}^\dagger \ t^{\dagger},$$
for all $t,s \in \mathcal{P}(V)$ and
leaving the relations in $\mathcal{P}(3)$ defining the operad $\mathcal{P}$ globally invariant (recall we suppose the notion $As^c-\mathcal{P}$-bialgebras holds so there is no relation between the generating operations of $\mathcal{P}(2)$).
This involution is the only anti-homomorphism of $\mathcal{P}$-algebras which agrees with the involution over $V$.
Extend this involution on $\mathcal{P}(V)^{\otimes 2}$ by the formula,
$$(t \otimes s)^{\dagger}:= s^{\dagger} \otimes t^{\dagger},$$
and operations as follows:
$$ (t \otimes s) \bullet_i (t' \otimes s'):=  (t \bullet_i t') \otimes (s \bullet_i s'),$$
for all $s,t \in \mathcal{P}(V)$ and $\bullet_i \in \mathcal{P}(2)$.
\begin{lemm}
\label{com-del}
Denote by $\delta$ the coassociative coproduct of the $As^c-\mathcal{P}$-bialgebra $\mathcal{P}(V)$.
Then, $\delta$ commutes with $\dagger$, that is:
$$ \delta(t^\dagger) = \delta(t)^\dagger, $$
for any $t \in \mathcal{P}(V)$.
\end{lemm}
\Proof
Observe that the involution $\dagger$ preserves the gradding.
Let $t \in V$. Then $t^\dagger \in V$ and $\delta(t)^{\dagger}=0=\delta(t^{\dagger})$.
Suppose $\delta(r)^{\dagger}=\delta(r^{\dagger})
$ for any element of degree up to $n$. Let $r$
be an element of degree $n+1$. As the operad $\mathcal{P}$ is binary, there exist $t$ and $s$ of smaller degrees and an operation $\bullet \in \mathcal{P}(2)$ such that
$r=t \bullet s$.
\begin{eqnarray*}
\delta(t \bullet s)^{\dagger} &=& s^{\dagger} \otimes t^{\dagger} + [s_{(2)}]^{\dagger} \otimes ([s_{(1)}]^{\dagger} \ {\bullet}^\dagger \ t^{\dagger}) + (s^{\dagger} \ {\bullet}^\dagger \ [t_{(2)}]^{\dagger} )\otimes [t_{(1)}]^{\dagger}, \\
&=& s^{\dagger} \otimes t^{\dagger} + \delta(s)^{\dagger} \ {\bullet}^\dagger \ t^{\dagger} + s^{\dagger} \ {\bullet}^\dagger \ \delta(t)^{\dagger},\\
&=& s^{\dagger} \otimes t^{\dagger} + \delta(s^{\dagger}) \ {\bullet}^\dagger \ t^{\dagger} + s^{\dagger} \ {\bullet}^\dagger \ \delta(t^{\dagger}), \ \textrm{(by induction)},\\
&=& \delta(s^{\dagger} \ {\bullet}^\dagger \ t^{\dagger}) = \delta((t \bullet s)^{\dagger}).
\end{eqnarray*}
\eproof

\begin{prop}
\label{com-del1}
Let $V$ be a $K$-vector space.
Let $(As, \mathcal{P}, Prim\mathcal{P})$ be a good triple of operads from Theorem~\ref{outil}.
Suppose the existence of an involution $\dagger$ on $V$ which extends to an involution $\dagger$ on $\mathcal{P}(V)$.
Then, $Prim\mathcal{P}(V)$ is invariant under $\dagger$.
\end{prop}
\Proof
Let $t$ be a primitive element. Then,
$\delta(t^{\dagger})=\delta(t)^{\dagger}=0$.
Hence the space of primitive elements of $Prim\mathcal{P}(V)$ is invariant under the involution $\dagger$.
\eproof

\noindent
We present a result linking right and left idempotents associated with an operation.
\begin{prop}
Let $V$ be a $K$-vector space.
The idempotents ${e^r}_{_\bullet},{e^l}_{_\bullet}: \
\mathcal{P}(V) \rightarrow Prim  \mathcal{P}(V),$ verify:
$$ \forall t \in \mathcal{P}(V), \ \  {e^r}_{_\bullet}(t^{\dagger})= {e^l}_{_\bullet^{\dagger}}(t)^{\dagger}, \ \ \ {e^l}_{_\bullet}(t^{\dagger})= {e^r}_{_\bullet^{\dagger}}(t)^{\dagger}.$$
\end{prop}
\Proof
We have ${e^r}_{_\bullet}(t^{\dagger})=t^{\dagger}={e^l}_{_\bullet}(t)^{\dagger},$ for any $t\in V$.
we proceed by induction on the degree and suppose the result holds for any element of degree equals at most $n$. As $\mathcal{P}$ is binary, any element of degree $n+1$ is of the form:
$t \star s$
with $t,s $ elements of smaller degrees and $\star \in \mathcal{P}(2)$.
On the one hand,
\begin{eqnarray*}
{e^r}_{_\bullet}((t \star s )^{\dagger}) &=& {e^r}_{_\bullet}(s ^{\dagger}\star^{\dagger} t^{\dagger}), \\
&=&
s ^{\dagger} \star^{\dagger} t^{\dagger} -[s ^{\dagger} \bullet {e^r}_{_\bullet}(t^{\dagger}) + [s ^{\dagger}]_{(1)}\bullet {e^r}_{_\bullet}([s ^{\dagger}]_{(2)} \star^{\dagger} t^{\dagger}) + (s ^{\dagger} \star^{\dagger} [t^{\dagger}]_{(1)}) \bullet {e^r}_{_\bullet}([t^{\dagger}]_{(2)})].
\end{eqnarray*}
On the other hand,
$$
{e^l}_{_\bullet^\dagger}(t \star s ) = t \star s -[{e^l}_{_\bullet^\dagger}(t) \bullet^\dagger s + {e^l}_{_\bullet^\dagger}(t_{(1)})\bullet^\dagger (t_{(2)} \star s) + {e^l}_{_\bullet^\dagger}(t \star s_{(1)}) \bullet^\dagger s_{(2)}].
$$
Therefore, taking the involution on both sides leads to:
$${e^l}_{_\bullet^\dagger}(t \star s )^\dagger = s^\dagger \star^\dagger t^\dagger -[s^\dagger \bullet {e^l}_{_\bullet^\dagger}(t)^\dagger +  (s^\dagger  \star^\dagger [t_{(2)}]^\dagger) \bullet {e^l}_{_\bullet^\dagger}(t_{(1)})^\dagger +  [s_{(2)}]^\dagger \bullet {e^l}_{_\bullet^\dagger}(t \star s_{(1)})^\dagger]. $$
Applying induction,
$${e^l}_{_{\bullet}^\dagger}(t \star s )^\dagger= s^\dagger \star^\dagger t^\dagger -[s^\dagger \bullet {e^r}_{_\bullet}(t^\dagger) +  (s^\dagger  \star^\dagger [t_{(2)}]^\dagger) \bullet {e^r}_{_\bullet}([t_{(1)}]^\dagger) +  [s_{(2)}]^\dagger \bullet {e^r}_{_\bullet}((t \star [s_{(1)}])^\dagger)]. $$
However, Lemma~\ref{com-del} shows that the decomposition of an element $r$ by $\delta$ leads to the following equalities: $[r_{(1)}]^\dagger=[r^\dagger]_{(2)}$ and $[r_{(2)}]^\dagger=[r^\dagger]_{(1)}$.
Therefore,
$${e^l}_{_\bullet^\dagger}(t \star s )^\dagger= s^\dagger \star^\dagger t^\dagger -[s^\dagger \bullet {e^r}_{_\bullet}(t^\dagger) +  (s^\dagger  \star^\dagger [t^\dagger]_{(1)}) \bullet {e^r}_{_\bullet}([t^\dagger]_{(2)}) +  [s^\dagger]_{(1)} \bullet {e^r}_{_\bullet}([s^\dagger]_{(2)} \star^\dagger t^\dagger)]. $$
Hence the first equality. For the second one, let $t$ be an element of $\mathcal{P}(V)$. Then, $${e^r}_{_\bullet^\dagger}(t)^\dagger= {e^r}_{_\bullet^\dagger}({t^\dagger} ^\dagger)^\dagger)= {e^l}_{_\bullet}(t^\dagger).$$
\eproof

\subsection{The structure theorem for L-algebras}
We now come back to the L-operad and obtain:
\begin{theo}
For any $As^c$-L-bialgebra, $\mathcal{H}$, the following are equivalent:
\begin{enumerate}
\item {The $As^c$-L-bialgebra $\mathcal{H}$ is connected.}
\item {There is an isomorphism of bialgebras $\mathcal{H} \simeq U(Prim \mathcal{H})$.}
\item{There is an isomorphism of connected coalgebras $\mathcal{H} \simeq As^c(Prim \mathcal{H})$.}
\end{enumerate}
\end{theo}

\noindent
Because of Theorem~\ref{outil}, we get an isomorphism of Schur functors $ \textrm{L} \simeq As^c \circ Prim \ \textrm{L}$. As explained in \cite{GB},
one can deduce the generating function of the operad $Prim \ \textrm{L}$. Indeed since the generating function of the operad $As^c$ is $f_{As^c}(x)=\frac{x}{1-x}$
and the generating function of the L-operad is  $f_{\textrm{L}}(x)=\frac{4}{3}\sin^2(\frac{1}{3}\textrm{asin}(\sqrt{\frac{27x}{4}})),$
we claim that the generating function of the operad $Prim \ \textrm{L}$ is,
$$ f_{_{Prim \ \textrm{L}}}(x)=\frac{\sin^2(\frac{1}{3}\textrm{asin}(\sqrt{\frac{27x}{4}}))}{\frac{3}{4} + \sin^2(\frac{1}{3}\textrm{asin}(\sqrt{\frac{27x}{4}}))}.$$
Its Taylor series starts with,
$$ f_{_{Prim \ \textrm{L}}}(x)= x+ x^2 + 4x^3 + 17x^4 + 81 x^5 + 412 x^6 + 2192 x^7 \ldots.$$
For instance,
$\dim \ Prim \ \textrm{L}_2= 1$ so $Prim \ \textrm{L}_2$ is spanned by $x \bowtie y:= x \succ y - x \prec y$ and  $\dim \ Prim \ \textrm{L}_3= 4$ so $Prim \ \textrm{L}_3$ is  spanned by,
$$[x,y,z]_1:=(x \bowtie y) \bowtie z; \ \ [x,y,z]_2:=x \bowtie (y \bowtie z); \ \ $$
$$[x,y,z]_3:=(x \succ y) \succ z - x \succ (y \succ z); \ \ [x,y,z]_4:=(x \prec y) \prec z - x \prec (y \prec z). \ \ $$
\section{Triplicial-algebras and L-algebras}
\label{Trip}
\subsection{A good triple of operads}
\label{good}
We call a triplicial-algebra a $K$-vector space equipped with $3$ operations verifying the following constraints:
$$ \forall \ 1\leq i\leq j \leq 3; \ \ (x \bullet_i y) \bullet_j z= x \bullet_i (y \bullet_j z).$$
In particular, all our operations are associative. We denote by \textsf{Trip} the corresponding category and by $Trip$ the associated operad. Note that $Trip$ is a binary quadratic regular and set-theoretic operad. Kill one of the three products to recover the definition of duplicial-algebras. The main interest of triplicial-algebras ($Trip$-algebras for short) lies in the following theorem.
\begin{theo}
The triple $(As, Trip, \textrm{L})$ is a good triple of operads. Therefore,
The category of connected $As^c-Trip$-bialgebras and the category of L-algebras are equivalent.
$$ \{\textrm{conn.} \ As^c-Trip-bialg.\} \underset{Prim}{\overset{U}{\leftrightarrows}} \{\textrm{L}-alg.\}$$
\end{theo}
\Proof
As $Trip$ is a regular quadratic binary operad with only entanglement relations, Theorem \ref{outil} claims that $(As, Trip, Prim \ Trip)$ is a good triple of operads.
Let $T$ be a $Trip$-algebra. The following operations,
$$x \succ y:= x \bullet_1 y - x \bullet_2 y, \ \ x \prec y:= x \bullet_3 y - x \bullet_2 y,$$
for all $x,y \in T$,
verify:
$$ (x \succ y) \prec z = x \succ (y \prec z),$$
and turns the $K$-vector space $T$ into a L-algebra. Moreover, if $T$ is a $As^c-Trip$-bialgebra then these operations are primitive operations.  Consequently, its primitive part is a L-algebra. Theorem \ref{asl} in the next section shows that $Trip(V)$ is isomorphic to $As(L(V))$ as triplicial-algebras. Since the coproduct in $As(L(V))$ is the usual deconcatenation and the triple $(As,As,Vect)$  endowed with the infinitesimal relation is good \cite{LodRon}, its primitive part is $L(V)$. Hence $Prim \ Trip=L$. Apply now \cite{GB}, Theorem 2.6.3. to conclude.
\eproof

\NB
This theorem is important since it allows to consider the (Markov) L-algebra associated with a given weighted directed graph as a connected $As^c-Trip$-bialgebra via the universal enveloping functor $U$, see Subsection~\ref{event}.

\subsection{Rota-Baxter maps and $Hom_K(T,T)$}
\label{Rota}
We give few words on the $K$-vector space, $Hom_K(T,T)$, of endomorphisms of a given $As^c-Trip$-bialgebra $(T, \bullet_1, \bullet_2, \bullet_3)$.
We introduce three convolution product defined as follows:
$$ f \bar{\bullet}_i g:= \bullet_i  (f \otimes g) \delta,$$
for all $f,g \in Hom_K(T,T)$ and $i=1,2,3$.
Then, observe that $(Hom_K(T,T), \bar{\bullet}_1, \bar{\bullet}_2,\bar{\bullet}_3)$ is also a $Trip$-algebra and for each $i=1,2,3$, $(Hom_K(T,T), \bar{\bullet}_i)$ can be endowed with a Ennea-algebra structure \cite{Lertribax}. There are at least 6 Rota-Baxter maps of weight 1 on $Hom_K(T,T)$ given by the shift operators:
$$ \beta_i (f):=id \bullet_i f, \ \ \gamma_i (f):= f \bullet_i id,$$
for all $f \in Hom_K(T,T)$ and $i=1,2,3$.
They obey the following commutation rules:
$$ \forall 1 \leq i \leq j \leq 3, \ \ \beta_i \circ \gamma_j = \gamma_j \circ \beta_i.$$
\subsection{Even trees and the triplicial-algebras}
\label{event}
An even tree of size $n$ is an ordered tree with $2n$ edges in which each node has an even output. Here are even trees of size 1 and 2.
\begin{center}
\includegraphics*[width=9cm]{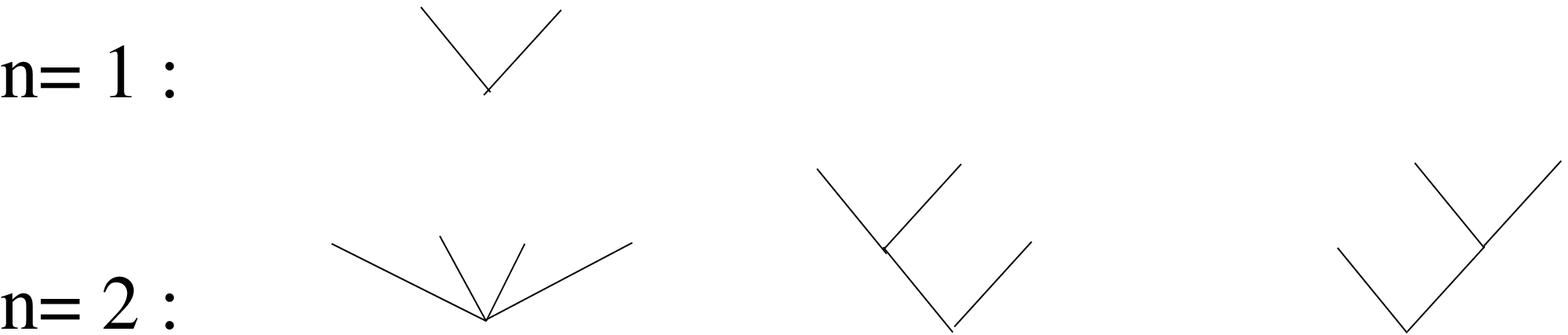}
\end{center}
The set of even trees of size $n$ is denoted by $\mathcal{E}_n$. It has been shown in \cite{DFN} that the cardinality of $\mathcal{E}_n$ is $\frac{1}{2n+1} {3n \choose n}$ which is also the number of planar rooted ternary trees on $n$ nodes, and also the number of symmetric planar rooted ternary trees on $2n$ nodes!

\noindent
Define the gluing operation for any positive integers $n,m$ as follows,
$$ \smile: \ \mathcal{E}_n \times \mathcal{E}_m \rightarrow \mathcal{E}_{n+m}, \ \ (t,s) \mapsto t \smile s,$$
where $t \smile s$ is the even tree of size $n+m$ obtained by gluing the root of $t$ to the root of $s$.
For instance $\treeAS \smile \treeAS = \treeCorr$.
Define also the following operations:
$$ \nearrow, \ \nwarrow: \ \mathcal{E}_n \times \mathcal{E}_m \rightarrow \mathcal{E}_{n+m},$$
as follows:
$t \nearrow s$ is the even tree of size $n+m$ obtained by setting $t$ on the most left leaf of $s$ and $t \nwarrow s$ is the even tree of size $n+m$ obtained by setting $s$ on the most right leaf of $s$.
Extend these three operations by $K$-bilinearity to get three binary operations,
$$\smile, \ \nearrow, \ \nwarrow: K\mathcal{E}_n \otimes K\mathcal{E}_m \rightarrow K\mathcal{E}_{n+m}.$$
\begin{theo}
Let $V$ be a $K$-vector space. Then, the $K$-vector space,
$$ \mathcal{E}ven(V):= \bigoplus_{n>0} \ K\mathcal{E}_{n} \otimes V^{\otimes n},$$
equipped with the three binary operations
still denoted by $\smile, \ \nearrow, \ \nwarrow$ and defined by:
$$ (t \otimes v_1 \otimes \ldots \otimes v_n) \bullet (s \otimes v'_1\otimes \ldots \otimes v'_m ):= (t \bullet s)\otimes v_1 \otimes \ldots \otimes v_n \otimes v'_1\otimes \ldots \otimes v'_m,$$
for $\bullet \in \{\smile, \ \nearrow, \ \nwarrow \}$ is the free $Trip$-algebra over $V$.
Its generating function is:
$$f_{Trip}(x)= \frac{2}{\sqrt{3x}}\sin(\frac{1}{3}\textrm{asin}(\sqrt{\frac{27x}{4}}))-1.$$
\end{theo}
\Proof
As the operad $Trip$ is regular, we need only to prove the theorem for a one dimensional vector space. Observe first that $\mathcal{E}ven(K)$ is a $Trip$-algebra. Using $\smile, \ \nearrow, \ \nwarrow$, one easily check that $K\mathcal{E}_2$ is generated by $\treeAS$. Suppose this is the case up to $K\mathcal{E}_n$. Let $t \in \mathcal{E}_{n+1}$. Then, suppose
there exist $k>1$ and $t_{i_1}, \ldots, t_{i_k}$, with $i_1 + \ldots + i_k=n$ and $t_{i_j} \in \mathcal{E}_{i_j}$ such that,
$$ t=t_{i_1} \smile \ldots \smile  t_{i_k}.$$
By induction, these even trees will be generated by $\treeAS$. If this is not possible, then $t$ is of the form $t_1 \nearrow \treeAS \nwarrow t_2$, or $t_1 \nearrow \treeAS$ or $\treeAS \nwarrow t_2$. Therefore by induction
$\mathcal{E}_{n+1}$ is generated by $\treeAS$. Let $(A, \bullet_1, \bullet_2,\bullet_3)$ be a $Trip$-algebra and $f: K \rightarrow A$, a map such that $f(1_K)=a$. Define $i: K \rightarrow \mathcal{E}ven(K), \ 1_K \mapsto \treeAS$ and $\tilde{f}: \mathcal{E}ven(K) \rightarrow A$ inductively as follows:
$$ \tilde{f}(\treeAS):= a, $$
$$ \tilde{f}(t \nearrow s):=\tilde{f}(t) \bullet_1 \tilde{f}(s), \ \tilde{f}(t \smile s):= \tilde{f}(t) \bullet_2 \tilde{f}(s), \ \tilde{f}(t \nwarrow s):=\tilde{f}(t) \bullet_3 \tilde{f}(s).$$
Then, $\tilde{f}$ is by construction the unique $Trip$-algebra morphism verifying $\tilde{f}\circ i =f$. Hence, $\mathcal{E}ven(K)$ is the free $Trip$ algebra on one generator.
As $\dim K\mathcal{E}_n = \frac{1}{2n+1} {3n \choose n}$, the generating function of the operad $Trip$ is the generating function of even trees or planar rooted ternary trees, hence the last claim.
\eproof

\NB \textbf{[The universal enveloping functor $U$]}
Recall notation of Subsection~\ref{good}. The functor $U$ acts as follows. Let $(L, \prec, \succ)$ be a L-algebra. Then $U(L)$ is given by $Trip(L)/ \sim$, where the equivalence relation $\sim$ consists in identifying,
$$ x \succ y:= x \nearrow y - x \smile y, \ \ x \prec y:= x \nwarrow y - x \smile y,$$
for all $x,y \in L$.

\subsection{Involutive triplicial-algebras}
\noindent
An involutive $Trip$-algebra $(T, \bullet_1, \bullet_2, \bullet_3)$ is a  triplicial-algebra equipped with an involution $\iota: T \rightarrow T$ verifying $\iota(x \bullet_1 y)=\iota(y) \bullet_3 \iota(x)$, $\iota(x \bullet_2 y)=\iota(y) \bullet_2 \iota(x)$, $\iota(x \bullet_3 y)=\iota(y) \bullet_1
\iota(x)$.
Observe that once $\bullet_1$ is given so is $\bullet_3$ and conversely.
The free $Trip$-algebra on one generator $\mathcal{E}ven(K)$ is an involutive $Trip$-algebra. Indeed, consider the involution over $\mathcal{E}ven(K)$ still denoted by $\dagger$ and defined inductively for any even trees $t,s$ by,
$$ \treeAS^{\dagger}=\treeAS, $$
$$ (s \nearrow t)^{\dagger}:=t^{\dagger} \nwarrow s^{\dagger}, \ (s \nwarrow t)^{\dagger}:=t^{\dagger} \nearrow s^{\dagger}, \ (s \smile t)^{\dagger}:=t^{\dagger} \smile s^{\dagger}.$$
An even tree is said to be symmetric if it is invariant under the involution $\dagger$.
Extend now this involution on $\mathcal{E}ven(K)^{\otimes 2}$ by the formula,
$$(t \otimes s)^{\dagger}:= s^{\dagger} \otimes t^{\dagger}.$$
\begin{theo}
The L-algebra of primitive elements of the $As^c-Trip$-bialgebra $\mathcal{E}ven(K)$ is invariant under the involution defined on even trees. Moreover, this involution coincides with the involution introduced on rooted planar ternary trees with odd degrees.
\end{theo}
\Proof
Use Proposition~\ref{com-del1} for the first part.
Recall that $Prim \ \mathcal{E}ven(K)$ is a L-algebra generated by $\treeAS$ and the two operations over even trees:
$$ t \succ s:= t \nearrow s - t \smile s, \ \ t \prec s:= t \nwarrow s - t \smile s.$$
Therefore,
$$ (t \succ s)^{\dagger}= (t \nearrow s)^{\dagger} - (t \smile s)^{\dagger}=s^{\dagger} \prec t^{\dagger} \ \ \textrm{and} \ \ (t \prec s)^{\dagger}= (t \nwarrow s)^{\dagger} - (t \smile s)^{\dagger}=s^{\dagger} \succ t^{\dagger}.$$
Identify $\treeAS$ to the word 1 coding the symmetric planar rooted tree on one node $\treeM$. Then the involution on even trees does coincide with the involution on rooted planar symmetric ternary trees we introduced in Subsection~\ref{L-alg} since both of the involution are defined recursively and agree with the generator 1.
\eproof

\noindent
Here is a way to construct other involutive triplicial-algebras.
Let $V$ be a $K$-vector space. Recall $As(V)$ is the free associative algebra over $V$. The concatenation will be used to denote the usual associative product in $As(V)$.
\begin{prop}
Let $(A, *)$ be an involutive associative algebra with involution denoted by $\iota:A \rightarrow A$, that is $\iota(a* a')=\iota(a')* \iota(a)$, for all $a,a' \in A$. Extend this involution on $A \otimes As(A)$ as follows,
$$ \iota(a \otimes a_1 \ldots a_n):= \iota(a_n)\otimes \iota(a_{n-1})\ldots \iota(a_1)\iota(a), $$
for any tensor $\omega:=a \otimes a_1 \ldots a_n \in A \otimes As(A)$ with the $a_i \in A$.
Then, $A\otimes As(A)$ equipped with the following operations,
$$ (a \otimes a_1  \ldots  a_n) \bullet_1 (a' \otimes  a'_1 \ldots  a'_m):=
[a* a'] \otimes a_1  \ldots  a_n  a'_1 \ldots  a'_m,$$
$$ (a \otimes a_1  \ldots a_n) \bullet_2 (a' \otimes a'_1 \ldots  a'_m):= a \otimes
a_1  \ldots  a_n  a' a'_1 \ldots a'_m, $$
$$ (a \otimes a_1  \ldots a_n) \bullet_3 (a' \otimes a'_1 \ldots  a'_m):= a \otimes
a_1  \ldots  a_{n-1}  a' a'_1 \ldots a'_{m-1}[a_n*a'_m], $$
where $a, a' \in A$ and $a_i, a'_i \in  As(A)$,
is an involutive $Trip$-algebra.
\end{prop}
\Proof
Straightforward.
\eproof

\NB
The $Trip$-monoid can be easily defined and its free $Trip$-monoid over a set $X$ is of course given by the even trees.\\
Diving out by the $Trip$-ideal generated by $t \smile s$ for all $t,s \in \mathcal{E}ven(K)$ gives the free duplicial-algebra on one generator $Dup(K)$ since only planar rooted binary trees will survive.\\
For another operad whose associated free object
is constructed over ternary trees the reader should read \cite{Ler-polyg}. We get also a coassociative coproduct over the free object but the infinitesimal relation linking the coproduct and binary operations has to be replaced by the so-called semi-Hopf relation.

\subsection{Another presentation of the free triplicial-algebra}
\noindent
We give here another presentation of the free triplicial-algebra over $V$ inspired by \cite{GB} Prop. 5.2.7.
\begin{prop}
\label{asl1}
Let $(L, \succ,\prec)$ be a L-algebra. Define on $As(L)$ the following operations:
$$(1) \ \ \ \  x \circ_1 y:= x \otimes y + x \succ y,$$
where the extension of the operation $\succ$, still denoted by $\succ$, is defined by induction as follows:
$$ (r_1): \ \ \ \ \ x \succ (y \otimes l'):=(x \succ y) \otimes l', $$
$$ (x \otimes l)\succ l'= -(x\succ l)\succ l'+ x\succ (l\succ l')+ x \otimes (l \succ l'),$$

$$(2) \ \ \ \ \ \ x \circ_2 y := x \otimes y,$$

$$(3) \ \ \ \ x \circ_3 y:= x \otimes y + x \prec y,$$
where the extension of the operation $\prec$, still denoted by $\prec$, is defined by induction as follows:
$$ (r_3): \ \ \ \ \ \ (l \otimes x) \prec y= l \otimes (x \prec y),$$
$$  l \prec (l' \otimes y)=(l \prec l')\prec y- l \prec(l' \prec x)+ (l \prec l')\otimes x, $$
for all $x,y \in As(L)$ and $l,l' \in L$.
Then, $As(L)$ is a triplicial-algebra.
\end{prop}
\Proof
Because of the construction of the extension of $\succ$ and $\prec$ and the relations $(r_1)$ and $(r_3)$, we have: $(x \otimes y) \prec z= x \otimes (y \prec z)$ and
$ x \succ (y \otimes z)= (x \succ y) \otimes z,$
for any $x,y,z \in As(L)$.
Therefore, we have the following equalities:
$ (x \circ_1 y)\circ_2 z= x \circ_1 ( y\circ_2 z),$ and
$ (x \circ_2 y)\circ_3 z= x \circ_2 ( y\circ_3 z),$
for any $x,y,z \in As(L)$.
We now establish the equality:
$$ E_1: \ \ \ \ (x \circ_1 y)\circ_3 z= x \circ_1 (y\circ_3 z).$$
Suppose degree of $y$ greater than 2.
Set $y:=y'\otimes l$, with $l \in L$. We have:
\begin{eqnarray*}
(x \circ_1 y)\circ_3 z &=& (x \circ_1 (y' \otimes l))\circ_3 z,\\
&= &   (x \circ_1 (y' \circ_2 l))\circ_3 z,\\
& =&  ((x \circ_1 y') \circ_2 l)\circ_3 z,\\
&= & (x \circ_1 y') \circ_2 (l \circ_3 z),\\
& =& x \circ_1 (y' \circ_2 (l \circ_3 z)),\\
& =& x \circ_1 ((y' \circ_2 l) \circ_3 z),\\
& = &x \circ_1 (y \circ_3 z),
\end{eqnarray*}
which proves that $(E_1)$ holds if degree of $y$ is greater than 2. We now fix degree of $y$ equal to 1 and set $y:=l' \in L$. If degree $x$ and $z$ are equal to 1, then $(E_1)$ holds because in a L-algebra
the relation $(l \succ l')\prec l''= l \succ (l' \prec l'')$ holds.
We suppose degre of $z$ greater than 2 and degree of $x$ equal to 1. We set $z = l\otimes z'$. On the one hand:
\begin{eqnarray*}
(x \circ_1 l') \circ_3 z &=& (x \circ_1 l') \circ_3  (l\otimes z'),\\
& =& (x \otimes l' + x \succ l') \circ_3  (l\otimes z'),\\
&=& x \otimes l' \otimes l\otimes z' + (x \otimes l')\prec (l\otimes z')+ (x \succ l') \otimes l\otimes z'+ (x \succ l')\prec (l\otimes z'),\\
&=& x \otimes l' \otimes l\otimes z' + x \otimes [ l'\prec (l\otimes z')]+ (x \succ l') \otimes l\otimes z'+ (x \succ l')\prec (l\otimes z'),\\
&=& x \otimes l' \otimes l\otimes z' \\
& & + x \otimes [ l'\prec (l\otimes z')]\\
& &+ (x \succ l') \otimes l\otimes z'\\
& & + ((x \succ l')\prec l)\prec z'- (x \succ l') \prec (l \prec z' ) + ((x \succ l')\prec l)\otimes z',\\
\end{eqnarray*}
On the other hand:
\begin{eqnarray*}
x \circ_1 (l' \circ_3 z) &=& x \circ_1 (l' \circ_3 (l \otimes z')),\\
& =&  x \circ_1 (l' \otimes l \otimes z'+ l' \prec ( l \otimes z')),\\
& =&  x \otimes l' \otimes l \otimes z'+ x \succ (l' \otimes l \otimes z') + x \otimes (l' \prec ( l \otimes z'))
+ x \succ (l' \prec ( l \otimes z')),\\
& =&  x \otimes l' \otimes l \otimes z'+ (x \succ l') \otimes l \otimes z' + x \otimes (l' \prec ( l \otimes z'))
+ x \succ (l' \prec ( l \otimes z')),\\
& =&  x \otimes l' \otimes l \otimes z' \\
& & + (x \succ l') \otimes l \otimes z' \\
& & + x \otimes (l' \prec ( l \otimes z'))\\
& & + x \succ [(l' \prec  l) \prec z' - l' \prec (l \prec z') + (l \prec l') \otimes z'],\\
& & + x \succ [(l' \prec  l) \prec z'] - x \succ [l' \prec (l \prec z')] + x \succ [(l \prec l')] \otimes z'],\\
\end{eqnarray*}
use now the relation $(l \succ l')\prec l''= l \succ (l' \prec l'')$ to conclude. We suppose now $(E_1)$ holds for any $x$ of degree lower than a fixed $n$ and take $x$ of degree $n+1$. Set $x= x' \otimes l$.
\begin{eqnarray*}
(x \circ_1 l') \circ_3 z &=&((x'\otimes l) \circ_1 l') \circ_3 z,\\
& =& (x'\otimes l \otimes l' + (x'\otimes l)\succ l') \circ_3 z,\\
& =& (x'\otimes l \otimes l' + x \succ (l\succ l') - (x \succ l)\succ l' + x \otimes (l\succ l')) \circ_3 z,\\
& =& x'\otimes l \otimes (l' \circ_3 z)\\
& & + (x \succ (l\succ l')) \circ_3 z \\
& & - ((x \succ l)\succ l')\circ_3 z\\
& & + (x \otimes (l\succ l')) \circ_3 z,\\
& =& x'\otimes l \otimes (l' \circ_3 z)\\
& & + (x \succ (l\succ l')) \circ_3 z \\
& & - ((x \succ l)\succ l')\circ_3 z\\
& & + (x \otimes (l\succ l')) \circ_3 z.
\end{eqnarray*}
We rewrite the last equation in terms of the triplicial operations.
\begin{eqnarray*}
(x \circ_1 l') \circ_3 z &= & x'\otimes l \otimes (l' \circ_3 z)
+ (x \succ (l\succ l')) \circ_3 z
 - ((x \succ l)\succ l')\circ_3 z
 + (x \otimes (l\succ l')) \circ_3 z  \\
&=& [x' \circ_2 (l \circ_1 l')]\circ_3 z \\
& & - [(x' \succ l) \circ_1 l']\circ_3 z\\
& & + [(x' \succ l) \circ_2 l']\circ_3 z\\
& & + [x' \circ_1 (l \succ l')]\circ_3 z\\
& & -[x' \circ_2 (l \succ l')]\circ_3 z,\\
&=& x' \circ_2 [(l \circ_1 l')\circ_3 z] \\
& & - (x' \succ l) \circ_1 [l'\circ_3 z], \ use \ induction \ and \ degree \ (x' \succ l)= degree \ x',\\
& & + (x' \succ l) \circ_2 [l'\circ_3 z]\\
& & + x' \circ_1 [(l \succ l')\circ_3 z],  \ use \ induction, \\\
& & -x' \circ_2 [(l \succ l')\circ_3 z],
\end{eqnarray*}
It is easy to show by induction that $(l \succ l')\circ_3 z= l \succ (l'\circ_3 z)$ whatever the degree of $z$ is. Therefore:
\begin{eqnarray*}
(x \circ_1 l') \circ_3 z &= &
x' \circ_2 [l \circ_1 (l'\circ_3 z)], \ use \ induction, \\
& & - (x' \succ l) \circ_1 [l'\circ_3 z]\\
& & + (x' \succ l) \circ_2 [l'\circ_3 z]\\
& & + x' \circ_1 [l \succ (l'\circ_3 z)], \\\
& & -x' \circ_2 [l \succ (l'\circ_3 z)],
\end{eqnarray*}
Use now the definitions of the triplicial operations and find:
\begin{eqnarray*}
(x \circ_1 l') \circ_3 z &= &
x'\otimes l \otimes (l' \circ_3 z)\\
& & + x \succ (l\succ (l' \circ_3 z)) \\
& & - (x \succ l)\succ (l'\circ_3 z)\\
& & + x \otimes (l\succ (l' \circ_3 z)),\\
& =& x'\otimes l \otimes (l' \circ_3 z) + (x'\otimes l) \succ (l' \circ_3 z), \ \ because  \ of \ the \ relation \ (r_1),\\
&= & (x'\otimes l) \circ_1 (l' \circ_3 z), \\
&= & x\circ_1 (l' \circ_3 z). \\
\end{eqnarray*}
Therefore $(E_1)$ holds.
For the associativity of $\circ_1$ and $\circ_3$,
we proceed again by induction.
Let $x,y,z \in L$.
\begin{eqnarray*}
(x \circ_3 y)\circ_3 z &=& (x \otimes y+ x \prec y)\circ_3 z,\\
&=& x \otimes y \otimes z + (x \otimes y) \prec z +
(x \prec y)\otimes z + (x \prec y)\prec z,\\
&=& x \otimes y \otimes z + x \otimes (y \prec z) +
(x \prec y)\otimes z + (x \prec y)\prec z,\\
\end{eqnarray*}

\begin{eqnarray*}
x \circ_3 (y\circ_3 z) &=& x \circ_3 (y \otimes z + y \prec z),\\
&=& x \otimes y \otimes z + x \otimes  (y \prec z) + x \prec (y \otimes z) + x \prec  (y \prec z),\\
&=& x \otimes y \otimes z + x \otimes  (y \prec z) + [(x \prec y) \prec z)- x\prec (y \prec z)+ (x\prec y) \otimes z ] \\
& & + x \prec  (y \prec z),\\
\end{eqnarray*}
hence the associativity of $\circ_3$ for any elements of degree 1. Suppose $x:=l$ an element of degree 1.
Without restriction on the degree of $y$ and $z$, we get:
\begin{eqnarray*}
(l \circ_3 y) \circ_3 z- l \circ_3 (y \circ_3 z) &=&
(l \prec y)\otimes z + (l \prec y)\prec z -l\prec(y \otimes z) -l\prec(y \prec  z).
\end{eqnarray*}
Let us show that:
$$ (E_2): \ \ \ \ \ \ (l \prec y)\otimes z + (l \prec y)\prec z = l\prec (y \otimes z) + l\prec (y \prec  z),$$
holds.
If the degree of $y$ is one without any assumptions on the degree of $z$ then $(E_2)$ holds.
We suppose $(E_2)$ holds for any $y$ of degree lower than a fixed $n$ and set $y:=l' \otimes y'$.
\begin{eqnarray*}
(l \prec y)\otimes z + (l \prec y)\prec z &=&
[(l\prec l') \prec y'] \otimes z -[l\prec (l' \prec y')]\otimes z + (l \prec l')\otimes y' \otimes z\\
& & + [(l\prec l')\prec y']\prec z - [l\prec (l'\prec y')]\prec z + [(l\prec l')\otimes y']\prec z,\\
&=&
[(l\prec l') \prec y'] \otimes z -[l\prec (l' \prec y')]\otimes z + (l \prec l')\otimes y' \otimes z\\
& & + [(l\prec l')\prec y']\prec z - [l\prec (l'\prec y')]\prec z + (l\prec l')\otimes (y'\prec z).
\end{eqnarray*}

\begin{eqnarray*}
l\prec (y \prec  z)+ l\prec (y \otimes z) &=& l\prec ((l' \otimes y') \prec z)
 + l \prec (l' \otimes y' \otimes z),\\
&=& l\prec (l' \otimes (y' \prec z))
 + l \prec (l' \otimes y' \otimes z),\\
&=& (l\prec l')\prec (y' \prec z) - l\prec(l' \prec (y' \prec z)) + (l\prec l')\otimes (y' \prec z)\\
& & + (l\prec l')\prec (y' \otimes z)- l\prec(l' \prec (y' \otimes z)) + (l\prec l')\otimes y' \otimes z.
\end{eqnarray*}
Gathering terms gives:
\begin{eqnarray*}
E_2 &:=& (l \prec y)\otimes z + (l \prec y)\prec z - l\prec (y \otimes z) - l\prec (y \prec  z),\\
&=& [(l\prec l')\prec y'] \otimes z + [(l\prec l')\prec y']\prec z - (l\prec l')\prec(y'\prec z) - (l\prec l')\prec(y' \otimes z)\\
& &-\{ [l \prec (l' \prec y')]\otimes z + [l\prec (l' \prec y')]\prec z - l\prec (l' \prec(y' \prec z)) -l \prec (l' \prec (y' \otimes z)).
\}
\end{eqnarray*}
The first row vanishes because of the induction hypothesis. The second row vanishes too because the degree of $(l' \prec y')$ is the same than the degree of $y'$. Apply now
the induction hypothesis twice:
\begin{eqnarray*}
[l \prec (l' \prec y')]\otimes z + [l\prec (l' \prec y')]\prec z &=& l \prec [(l' \prec y')\otimes z] + l\prec [(l' \prec y')\prec z], \ use \ induction, \\
&=& l \prec [(l' \prec y')\otimes z + (l' \prec y')\prec z],\\
&=& l \prec [l' \prec (y'\otimes z) + l' \prec (y'\prec z)], \ use \ again \ induction.
\end{eqnarray*}
Therefore, if degree of $x$ is one, the following,
$$ (x \circ_3 y)\circ_3 z=x \circ_3 (y\circ_3 z)$$
holds for any $y,z \in As(L)$.
Suppose $x$ of degree greater than 2. Set $x:=x'\otimes l$, $l \in L$. We get:
\begin{eqnarray*}
(x \circ_3 y)\circ_3 z &=& ((x'\otimes l) \circ_3 y)\circ_3 z,\\
&=& ((x'\circ_2 l) \circ_3 y)\circ_3 z,\\
&=& (x'\circ_2 (l \circ_3 y))\circ_3 z,\\
&=& x'\circ_2 ((l \circ_3 y)\circ_3 z),\\
&=& x'\circ_2 (l \circ_3 (y\circ_3 z)), \ \ since \ degree \ l = 1,\\
&=& (x'\circ_2 l) \circ_3 (y\circ_3 z),\\
&=& x \circ_3 (y\circ_3 z),
\end{eqnarray*}
hence the associativite of $\circ_3$. For $\circ_1$ we proceed similarly.
As $\circ_2$ is associative, $As(L)$ is a triplicial-algebra.
\eproof

\begin{theo}
\label{asl}
Let $V$ be a $K$-vector space. Then $As(L(V))$ is the free triplicial-algebra over $V$.
\end{theo}
\Proof
The generating function of the operad $Trip$ is given by:
$$f_{Trip}(x)= \frac{2}{\sqrt{3x}}\sin(\frac{1}{3}\textrm{asin}(\sqrt{\frac{27x}{4}}))-1.$$
As the generating function of the operad $As$ is $f_{As}(x):=\frac{x}{1-x}$, the computation
$f_{As}^{-1}\circ f_{Trip}$ gives the generating function of the L-operad. Indeed, dealing with the generating functions of operads, we know from Koszulity of the L-operad that $f_{L!}(-f_L(-x))=x$. As, $f_{L!}(x)=x^3 +2x^2 +x=x(x+1)^2$, we get:
$$ f_L(x)(f_L(x)-1)^2=x, \ \textrm{that is,} \ \ \ \frac{f_L(x)}{x}=\frac{1}{(f_L(x)-1)^2}.$$
But $f_{Trip}(x)=
\sqrt{\frac{f_L(x)}{x}}-1$ leads to,
$$f_{As}^{-1}\circ f_{Trip}(x)=\frac{\sqrt{\frac{f_L(x)}{x}}-1}{\sqrt{\frac{f_L(x)}{x}}}.$$
Therefore, $f_{As}^{-1}\circ f_{Trip}(x)= f_L(x) $ and for all $n>0$, $\dim \ Trip(n) = \dim \ (As\circ \textrm{L})(n)$. Let $V$ be a $K$-vector space. The usual inclusion map $V \hookrightarrow As(L(V))$ induces a unique triplicial-morphism $Trip(V)\rightarrow As(L(V))$ which turns out to be surjective by construction. As $ f_{Trip}= f_{As}\circ f_L$, this map is an isomorphism. Hence the claim.
\eproof

\NB
This theorem allows to code even trees via forests of symmetric ternary trees in a bijective way or with the help of words made out with $1$ and $\bar{2}$.

\subsection{Commutative $Trip$-algebras}
\label{comtrip}
Let $T$ be a $Trip$-algebra. Define new operations $\bullet_1^{op}, \ \bullet_2^{op}, \ \bullet_3^{op}$ by,
$$ x \bullet_1^{op} y := y \bullet_3 x, \
x \bullet_2^{op} y := y \bullet_2 x, \ x \bullet_3^{op} y := y \bullet_1 x, $$
for all $x,y \in T$. The $K$-vector space $T$ equipped with these three new operations is a new $Trip$-algebra, called the opposite $Trip$-algebra denoted by $T^{op}$. A $Trip$-algebra is said to be commutative when it coincides with its opposite structure. Therefore, a commutative $Trip$-algebra $T$ is a $K$-vector space equipped with two associative binary operations $\bullet$ and $\star$, verifying:
$$ x \bullet y = y \bullet x;$$
$$ (x \bullet y) \star z = x \bullet (y \star z);$$
$$ x \star y \star z = x \star z \star y.$$
Observe that $(T,\bullet)$ is a commutative associative algebra and that $(T,\star)$ is a permutative algebra. The operad $Perm$ of  permutative algebras is the Koszul dual of the right Pre-Lie operad and was introduced in \cite{Chp1}. The operad of commutative algebras are denoted by $ComTrip$. Observe that there exit three canonical functors:
$$ ComTrip \rightarrow Com, \ ComTrip \rightarrow Perm \rightarrow As, \ ComTrip \rightarrow \textrm{L}.$$

\noindent
Let $V$ be a $K$-vector space. Denote by $Com(V)$ the free associative commutative algebra over $V$ and by $UCom(V) = K \oplus Com(V),$ its augmented version. Denote by $\vee$ its usual commutative associative product.

\begin{prop}
Let $(A, *)$ be a commutative associative algebra.
Then, $A\otimes UCom(A)$ equipped with the following operations,
$$ (a \otimes a_1 \vee \ldots \vee a_n) \bullet_1 (a' \otimes  a'_1 \vee \ldots  \vee a'_m):=
[a* a'] \otimes a_1 \vee \ldots \vee a_n  \vee a'_1 \ldots  \vee a'_m,$$
$$ (a \otimes a_1 \vee  \ldots \vee a_n) \bullet_2 (a' \otimes a'_1 \vee \ldots  \vee a'_m):= a \otimes
a_1 \vee \ldots  a_n  \vee a' \vee a'_1 \vee \ldots \vee a'_m. $$
where $a, a' \in A$ and $a_i, a'_i \in  UCom(A)$,
is a commutative $Trip$-algebra.
\end{prop}
\Proof
Straightforward.
\eproof

\begin{theo}
The free commutative $Trip$-algebra over $V$ is the $K$-vector space:
$$ ComTrip(V):= Com(V)\otimes UCom(Com(V)),$$
equipped with the following binary operations:
$$ (\omega \otimes \omega_1 \vee \ldots \vee \omega_n) \vee (\xi \otimes \xi_1\vee \ldots \vee \xi_m):= \omega \otimes
\omega_1 \vee \ldots \vee \omega_n \vee \xi\vee \xi_1\vee \ldots \vee \xi_m. $$
$$ (\omega \otimes \omega_1 \vee \ldots \vee \omega_n) \bullet (\xi \otimes  \xi_1\vee \ldots \vee \xi_m):=
[\omega \vee \xi] \otimes \omega_1 \vee \ldots \vee \omega_n \vee \xi_1\vee \ldots \vee \xi_m,$$
where $\omega, \xi \in Com(V)$ and $\omega_i, \xi_i \in UCom(V)$.
Its generating function is:
$$ f_{_{ComTrip}}(x)= (exp(x)-1)exp(exp(x)-1)=x + 3\frac{x^2}{2!} + 10\frac{x^3}{3!} + 37\frac{x^4}{4!}+ 151\frac{x^5}{5!}+ \cdots. $$
\end{theo}
\Proof
Observe that equipped with such operations, $ComTrip(V)$ is a commutative $Trip$-algebra. Let $i: V \hookrightarrow V \otimes K \hookrightarrow Com(V)\otimes UCom(Com(V))$ be the usual inclusion map defined by $i(v):=v \otimes 1_K$. Let $(A, \bullet, \star)$ be another commutative $Trip$-algebra and $f:V \rightarrow A$ a linear map.
Define $\tilde{f}: ComTrip(V) \rightarrow A$ by induction as follows:
$$\tilde{f}(\omega \otimes \omega_1 \vee \ldots \vee \omega_n):= \tilde{f}(\omega) \star \tilde{f}(\omega_1) \star \ldots \star \tilde{f}(\omega_n),$$
and if $\omega:=v_1 \vee \ldots \vee v_m$,
$$\tilde{f}(\omega \otimes 1_K):=\tilde{f}(v_1 \vee \ldots \vee v_m):= f(v_1) \bullet \ldots \bullet f(v_m).$$
If $X$ and $Y$ are monomials of $ComTrip(V)$, the relation $\tilde{f}(X \vee Y)=\tilde{f}(X) \star \tilde{f}(Y)$ holds by construction.
If $X:= \omega \otimes 1_K$ and $Y:=\xi \otimes 1_K$, then
$\tilde{f}(X \bullet Y)= \tilde{f}([\omega \vee \xi] \otimes 1_K)= \tilde{f}(X) \bullet \tilde{f}(Y)$ by construction. This will be helpful in the next computation.
Set now $X:= \omega \otimes X'$ and $Y:=\xi \otimes Y'$, where $X'$ and $Y'$ are monomials of $Com(Com(V))$.
We get:\\
$\tilde{f}(X) \bullet \tilde{f}(Y)=(\tilde{f}(\omega) \star \tilde{f}(X')) \bullet (\tilde{f}(\xi) \star \tilde{f}(Y'))= [(\tilde{f}(\omega) \star \tilde{f}(X')) \bullet \tilde{f}(\xi)] \star \tilde{f}(Y')=[ \tilde{f}(\xi)\bullet(\tilde{f}(\omega) \star \tilde{f}(X'))] \star \tilde{f}(Y')= [ (\tilde{f}(\xi) \bullet \tilde{f}(\omega)) \star \tilde{f}(X')] \star \tilde{f}(Y')= [\tilde{f}(\xi \bullet \omega) \star \tilde{f}(X')] \star \tilde{f}(Y')= [\tilde{f}(\omega \bullet \xi) \star \tilde{f}(X')] \star \tilde{f}(Y')=\tilde{f}(\omega \bullet \xi) \star \tilde{f}(X') \star \tilde{f}(Y')= \tilde{f}([\omega \bullet \xi] \otimes X' \vee Y')=\tilde{f}(X \bullet Y).
$
We use twice the entanglement relation between $\bullet$ and $\star$ and the fact that $\bullet$ is commutative. We have shown that $\tilde{f}$ is the only $ComTrip$-algebra morphism which extends $f$. Therefore, $ComTrip(V)$ is the free $K$-vector space over $V$. The last claim is just a result concerning composition of functors (recall that $f_{Com}(x):=exp(x)-1$).
\eproof

\NB
Duplicial-algebras also admit an opposite structure and we can deals with commutative
duplicial-algebras. They are $K$-vector spaces equipped with an associative binary operation which verify:
$$ x yz= xzy.$$
Hence,
$$ComDup=Perm=ComDias,$$
where $ComDias$ is the operad of commutative dialgebras \cite{Loday}.
\NB
The previous theorem suggests the existence of a triple of operads $(Perm, ComTrip, Com)$ or $(NAP, ComTrip, Com)$. Does they exist?
\subsection{Dual of the $Trip$-operad and triangular numbers}
\label{dualtrip}
We go on our knowledge of $Trip$-algebras and present here its dual. A quasi-nilpotent $Trip$-algebra, $QNTrip$-algebra for short, $Q$ is $K$-vector space equipped with three binary operations $\bullet_i$, $i=1,2,3,$ verifying:
$$ \forall 1 \leq i \leq j \leq 3, \ \ (x \bullet_i y)\bullet_j z=  x \bullet_i ( y\bullet_j z),$$
$$ i>j \Rightarrow (x \bullet_i y) \bullet_j z =0,$$
$$ i>j \Rightarrow x \bullet_i (y \bullet_j z) =0,$$
for all $x,y,z \in  Q$. Therefore, there is a functor $\textsf{QNTrip} \rightarrow \textsf{Trip}$.
\begin{theo}
Let $V$ be a $K$-vector space and $UAs(V)$ the augmented free associative algebra. Consider the $K$-vector space
$$ QNTRip(V):= UAs(V)\otimes [V \otimes UAs(V)] \otimes UAs(V),$$
equipped with the following three binary operations,
$$(\omega \otimes [v \otimes \xi] \otimes \theta) \bullet_1  (\omega' \otimes [v' \otimes \xi'] \otimes \theta') = \Upsilon(\xi \otimes \theta) \ (\omega v \omega') \otimes [v' \otimes \xi'] \otimes \theta',$$
$$(\omega \otimes [v \otimes \xi] \otimes \theta) \bullet_2  (\omega' \otimes [v' \otimes \xi'] \otimes \theta') = \Upsilon(\theta \otimes \omega') \ \omega \otimes [v \otimes (\xi  v'  \xi')] \otimes \theta',$$
$$(\omega \otimes [v \otimes \xi] \otimes \theta) \bullet_3  (\omega' \otimes [v' \otimes \xi'] \otimes \theta') = \Upsilon(\omega'\otimes \xi') \ \omega \otimes [v \otimes \xi] \otimes (\theta v \theta'),$$
where Greek letters denote elements from $UAs(V)$, $v,v' \in V$ and $\Upsilon: UAs(V) \otimes UAs(V) \rightarrow K$ is the canonical projection map. Then, $QNTRip(V)$
is the free $ QNTRip$-algebra over $V$.
The dimension of $QNTRip_n$ is the triangular number
$\frac{n(n+1)}{2}$ and
its generating function is,
$$ f_{QNTrip}(x)=\frac{x}{(1-x)^3}= \sum_{n>0} \ \frac{n(n+1)}{2} x^n.$$
\end{theo}
\Proof
The $K$-vector space $QNTRip(V)$ equipped with these three operations is a $QNTrip$-algebra. Let $i: V \hookrightarrow K \otimes (V \otimes K) \otimes K$ be the expected inclusion map. Let $(A, \bullet_1, \bullet_2, \bullet_3)$ be another $QNTrip$-algebra and $f:V \rightarrow A$ be a linear map. Define $\tilde{f}: QNTRip(V) \rightarrow A$ inductively as follows:
$$ \tilde{f}(1_K \otimes [v \otimes 1_K] \otimes 1_K]:=f(v);$$
$$\tilde{f}(\omega \otimes [v \otimes \xi] \otimes \theta):= \tilde{f}(\omega) \bullet_1 f(v) \bullet_2  \tilde{f}(\xi) \bullet_3 \tilde{f}(\theta), $$
$$ \tilde{f}(v_1 \otimes \ldots \otimes  v_n):=f(v_1)\bullet_k \ldots \bullet_k f(v_n),$$
where $k=1,2,3$ respectively if the monomial $\rho:=
v_1 \otimes \ldots \otimes  v_n$ belongs to the first, second or third copy of $As(V)$. We understand this definition as follows: If $\omega=1_K$ (resp. $\xi =1_K$, resp. $\theta =1_K$) then $\tilde{f}(\omega) \bullet_1 $ (resp. $\bullet_2  \tilde{f}(\xi)$, resp. $\bullet_3 \tilde{f}(\theta)$)
vanishes in the right hand side of the middle equation.
It is not hard to see that $\tilde{f}$ so defined is the unique $QNTrip$-algebra morphism extending $f$. Therefore, $QNTRip(V)$ is the free $QNTRip$-algebra over $V$.
For the last claim, recall that the generating function of the operad $UAs$ is $f_{UAs}(x)=\frac{1}{1-x}$.
\eproof

\section{Anticyclic operads and invariant bilinear maps}
\label{anticyclic}
Anti-cyclic operads allow to deal with invariant antisymmetric bilinear maps.
The reader should read \cite{GeK, Markl} for the theory and \cite{Chp} for examples. We show that such maps cannot exist on L-algebras but do exist on the operad $[n]-Mag$, $(n >1)$.
Let $\mathcal{P}$ be a regular binary and quadratic operad and $A$ be a $\mathcal{P}$-algebra.
An invariant antisymmetric bilinear map on $A$ with values in some vector space $V$, $\bra \cdot \ ; \ \cdot \ket : A^{\otimes 2} \rightarrow V$ is by definition,
\begin{enumerate}
 \item {A collection of a map $\tau_n: \mathcal{P}_n \rightarrow \mathcal{P}_n$, $n \geq 2$ of order $n+1$ verifying,
$$ \bra \gamma(x_{1}, \ldots, x_{n}) ; \ x_{n+1} \ket = \bra \tau_n(\gamma)(x_{2}, \ldots, x_{n+1}) ; \ x_{1} \ket,$$
for any generating operation $\gamma \in \mathcal{P}_n$,}
\item{And a map $\tau_1: (\mathcal{P}_1 \simeq K) \rightarrow K$ defined by $1_K \mapsto -1_K$.}
\end{enumerate}

\begin{theo}
Let $A$ be a $L$-algebra.
There exists no invariant antisymmetric bilinear map on $A$.
\end{theo}
\Proof
Suppose there exists an invariant antisymmetric bilinear map on $A$. Then, there exists a map $\tau_2:\mathcal{L}_2 \rightarrow \mathcal{L}_2$ of order 3. As $\mathcal{L}_2=K\{ \prec \} \oplus K\{\succ\}$, $\tau_2$ is a two by two matrix written in this basis as,
$$ \tau_2 = \begin{pmatrix}
 a & b\\
c & d
\end{pmatrix}.
$$
As the L-operad is binary, the maps $\tau_n$ for $n>2$, if they exist, are built from $\tau_2$ (and thus are unique).
We now show that such a $\tau_2$ does not exist.
We compute,
\begin{eqnarray*}
0= \bra x\succ ( y\prec z) - (x\succ y)\prec z ; \ t \ket &=&
\bra \tau_2(\succ)[ ( y\prec z),t] ; x \ket - \bra \tau_2(\prec)[z,t]; (x\succ y) \ket,\\
&=&
\bra c( y\prec z)\prec t + d( y\prec z)\succ t  ; x \ket + \bra (x\succ y); \tau_2(\prec)[z,t]  \ket,\\
&=& \bra c( y\prec z)\prec t + d( y\prec z)\succ t  ; x \ket + \bra  \tau_2(\succ)[y; \tau_2(\prec)[z,t]]; x  \ket,\\
&=& \bra c( y\prec z)\prec t + d( y\prec z)\succ t + ac \ y\prec(z\prec t)  \\
& & +bc \ y\prec(z\succ t)
+ad \ y\succ (z\prec t) + bd \ y\succ (z\succ t) ;x  \ket.
\end{eqnarray*}
Hence, $c$ and $d$ have to vanish. Now, $\tau_2$ cannot be of order 3.
\eproof

\NB
The same proof holds for $Trip$-algebras.
Let us focus on the operad $[n]-Mag$, $n>1$. Set,
$$ C = \begin{pmatrix}
 0 & -1\\
1 & -1
\end{pmatrix},
$$
be the matrix used in \cite{Chp}. This matrix is of order 3.
If $n:=2p$, then construct $\tau_2$ as a diagonal block of $p$ matrices $C$ or its transpose. If $n:=2(p-1) +3$, then construct $\tau_2$ as a diagonal block of $p-1$ matrices $C$ or its transpose and the following matrix used one time:
$$
Perm = \begin{pmatrix}
 0 & 1 & 0\\
0 & 0 & 1 \\
1 & 0 & 0
\end{pmatrix}.
$$
The map $\tau_2: [n]-Mag_2 \rightarrow [n]-Mag_2 $  is thus of order three. As the operad $[n]-Mag$ is binary, the other maps $\tau_n$, $n>2$, will be uniquely determined by $\tau_1$ and $\tau_2$.
One could have also used the following map,
$$\tau_2 = \begin{pmatrix}
 C & 0\\
0 & Id_{n-2}
\end{pmatrix}.
$$
Hence, the following holds.
\begin{theo}
For a fixed $\tau_2$, there exits a unique collection of maps $\tau_n, \ n>2$, extending $\tau_1$ and $\tau_2$ and turning the regular operad $[n]-Mag$, $n>1$, into an anticyclic operad.
\end{theo}
\NB
For the case $n=1$, so is the operad $[1]-Mag:=Mag$, if we assume the existence of an element $j \in K$ of order three (take for instance $K=\mathbb{C}$ and a solution of $1 + x + x^2=0$.).

\noindent
\textbf{Acknowledgments:}
The author is indebted to E. Deutsch for sending him papers \cite{AMS-D, DFN} and to J.-L. Loday for his comments and for improving Proposition~\ref{asso-GK}.
\bibliographystyle{plain}
\bibliography{These}

\end{document}